\documentclass{aa} 

\usepackage{graphicx}
\usepackage{natbib}
\usepackage{longtable}
\usepackage{rotating}
\usepackage{lscape}
\usepackage{longtable}
\usepackage{amsmath}
\usepackage{overpic,color}
\usepackage{url}
\usepackage{hyperref}
\usepackage{breakurl}
\usepackage{xcolor}
\usepackage{txfonts}

\def \xmm {\emph{XMM-Newton}}

\def \sw {{\it Swift}}
\def \chandra {{\it Chandra}}

\def \Gal {\mbox{{NGC~1512}}}
\def \system {\mbox{{NGC~1512/1510}}}

\begin{document}
   \title{\emph{XMM-Newton} observation of the interacting galaxies\\ NGC~1512 and NGC~1510
          \thanks{Based on 
            observations obtained with \emph{XMM-Newton}, an ESA science mission with 
            instruments and contributions directly funded by ESA Member States 
            and NASA. The radio observations were obtained with 
              the Australia Telescope Compact Array, which is part of the Australia Telescope National
              Facility funded by the Commonwealth of Australia for operation as a
              National Facility managed by CSIRO.}\fnmsep
   \thanks{Tables~\ref{Tab. source list} and~\ref{Tab. source list classification} 
    are only available in electronic form
    at the CDS via anonymous ftp to cdsarc.u-strasbg.fr (130.79.128.5)
    or via http://cdsweb.u-strasbg.fr/cgi-bin/qcat?J/A+A/ }
   }

   \subtitle{}

   \author{L. Ducci
          \inst{1,2}
          \and
          P. J. Kavanagh\inst{1}
          \and
          M. Sasaki\inst{1}
          \and
          B. S. Koribalski\inst{3}
          }
   \institute{Institut f\"ur Astronomie und Astrophysik, Eberhard Karls Universit\"at, 
              Sand 1, 72076 T\"ubingen, Germany\\
              \email{ducci@astro.uni-tuebingen.de}
              \and
              ISDC Data Center for Astrophysics, Universit\'e de Gen\`eve, 16 chemin d'\'Ecogia, 1290 Versoix, Switzerland
              \and
              Australia Telescope National Facility, CSIRO Astronomy and Space Science, PO Box 76, Epping, NSW 1710, Australia
             }
   \date{}

  \abstract
{The galaxy NGC~1512 is interacting with the smaller galaxy NGC~1510 and shows a peculiar morphology,
characterised by two extended arms immersed in an HI disc whose size is about four
times larger than the optical diameter of NGC~1512.}
{For the first time we performed a deep X-ray observation of the galaxies NGC~1512 and NGC~1510 
with \emph{XMM-Newton} to gain information on the
population of X-ray sources and diffuse emission in a system of interacting galaxies.}
{We identified and classified the sources detected in the \xmm\ field of view
by means of spectral analysis, hardness-ratios calculated with a Bayesian method,
X-ray variability, and cross-correlations with catalogues in optical, infrared, and radio wavelengths.
We also made use of archival \sw\ (X-ray) and Australia Telescope Compact Array (radio) 
data to better constrain the nature of the sources detected with \xmm.}
{We detected 106 sources in the energy range of 0.2$-$12~keV, out of which 15 
are located within the $D_{25}$ regions of NGC~1512
and NGC~1510 and at least six sources coincide with the extended arms.
We identified and classified six background objects and six foreground stars.
We discussed the nature of a source within the $D_{25}$ ellipse of NGC~1512,
whose properties indicate a quasi-stellar object or an intermediate ultra-luminous X-ray source.
Taking into account the contribution of low-mass X-ray binaries and active galactic nuclei,
the number of high-mass X-ray binaries detected within the $D_{25}$ region of NGC~1512
is consistent with the star formation rate obtained in previous works based on 
radio, infrared optical, and UV wavelengths.
We detected diffuse X-ray emission from the interior region of \Gal\ with a plasma temperature of
$kT=0.68~(0.31-0.87)$~keV and a 0.3--10~keV X-ray luminosity of $1.3 \times 10^{38}$~erg~s$^{-1}$, 
after correcting for unresolved discrete sources.}
{}

   \keywords{galaxies: individual; NGC\,1512, NGC\,1510 $-$ X-rays: galaxies}

   \maketitle
%

\section{Introduction}
\label{section introduction}

NGC~1512 and NGC~1510 are two interacting galaxies
at a distance of $\sim$9.5~Mpc \citep{Koribalski04}.
NGC 1512 is a barred spiral galaxy 
(SB(r)a; \citealt{deVaucouleurs91}) with two star-forming rings
and two extended arms immersed in an HI disc about four times
larger than the optical diameter, probably caused by an
ongoing tidal interaction with the neighbouring galaxy NGC~1510 
\citep{Kinman78}.
NGC~1510 is a blue compact dwarf galaxy separated by only $\sim$14~kpc
from NGC~1512. Its blue colour and emission line spectrum are probably caused by
the star formation activity produced by the interaction with NGC~1512 \citep{Hawarden79}.
Based on the analysis of the distribution and kinematics of the HI gas 
and the star formation activity in NGC~1512/1510, \citet{Koribalski09} 
concluded that the star formation activity in the outskirts of the disc
as well as the distortion in the HI arms are the consequence of 
the interaction between the two galaxies that started about 400~Myr ago \citep{Koribalski09}.
These authors used methods based on measurements of lines and continuum emission
at different wavelengths (from radio to UV) to estimate the
star formation rate (SFR) of both NGC~1512 and NGC~1510.
They derived an average SFR of $0.22$~$M_\odot$~yr$^{-1}$ for NGC~1512
and $SFR=0.07$~$M_\odot$~yr$^{-1}$ for NGC~1510.

To our knowledge, there has been no detection of sources
at the position of NGC~1512/1510 with X-ray telescopes to date.
From ROSAT/PSPC observations, \citet{OSullivan01} derived
a luminosity upper limit for NGC 1510 of 
$L_{\rm x} \leq 5.8 \times 10^{39}$ erg s$^{-1}$ ($0.1-2.4$ keV).
Therefore, the work described in this paper represents the first report
of significant X-ray emission in NGC1512/1510.
The \xmm\ observation on which this paper is based (see Sect. \ref{sect. data analysis})
allows the detection of the brightest populations of X-ray sources ($L_x \gtrsim 10^{37}$ erg s$^{-1}$)
typically observed in normal galaxies.
The main class of galactic X-ray sources 
is that of X-ray binaries (XRBs), whose X-ray luminosity,
ranging from $\sim10^{32}$ to $\sim10^{39}$~erg~s$^{-1}$,
is produced by the accretion of matter from the donor star to a compact object (a neutron star or a black hole).
XRBs are conventionally divided into two classes according to the nature of the donor star:
low-mass X-ray binaries (LMXBs) and high-mass X-ray binaries (HMXBs).
LMXBs host late-type stars and have a lifetime of $\sim 10^7-10^9$~yr
because of the nuclear timescale of the donor star.
Their number is roughly proportional to the total stellar mass of the galaxy (e.g. \citealt{Gilfanov04}).
HMXBs host early-type stars and have a lifetime of $\sim 10^6-10^7$~yr.
Therefore, their number is proportional to the recent SFR of a galaxy (e.g. \citealt{Grimm03}).

Large-scale diffuse X-ray emission in galaxies can be used to trace outflows and feedback processes 
on galactic scales. Such emission has been detected with the current generation of 
X-ray observatories in many nearby galaxies, including M31 \citep{Shirey2001}, 
M101 \citep{Kuntz2003}, NGC~300 \citep{Carpano2005}, NGC~6946 \citep{Schlegel2003}, 
NGC~1672 \citep{Jenkins2011}, and NGC~253 \citep{Bauer08}. As such, the \system\ system may also exhibit 
diffuse X-ray emission that can be used to investigate these physical processes.

In this paper we report the results obtained from a study of the X-ray
sources in the \xmm\ field of view of NGC~1512/1510. The paper is organised as follows:
in Sect. \ref{sect. data analysis} we present the \xmm, \sw, 
and Australia Telescope Compact Array (ATCA) observations on which
our work is based and the analysis of the data.
In Sect. \ref{sect. analysis} we describe the methods adopted
to classify the sources based on their X-ray properties
and the analysis of X-ray diffuse emission from NGC~1512.
In Sect. \ref{sect. source classification} we cross-correlate the list of sources detected
with \xmm\ with radio, infrared, and optical catalogues
and discuss the identifications and classifications obtained for some of these sources.
In Sect. \ref{sect. discussion} the results for point sources and X-ray diffuse emission are discussed.

\begin{figure*}
\begin{center}
\includegraphics[width=10cm]{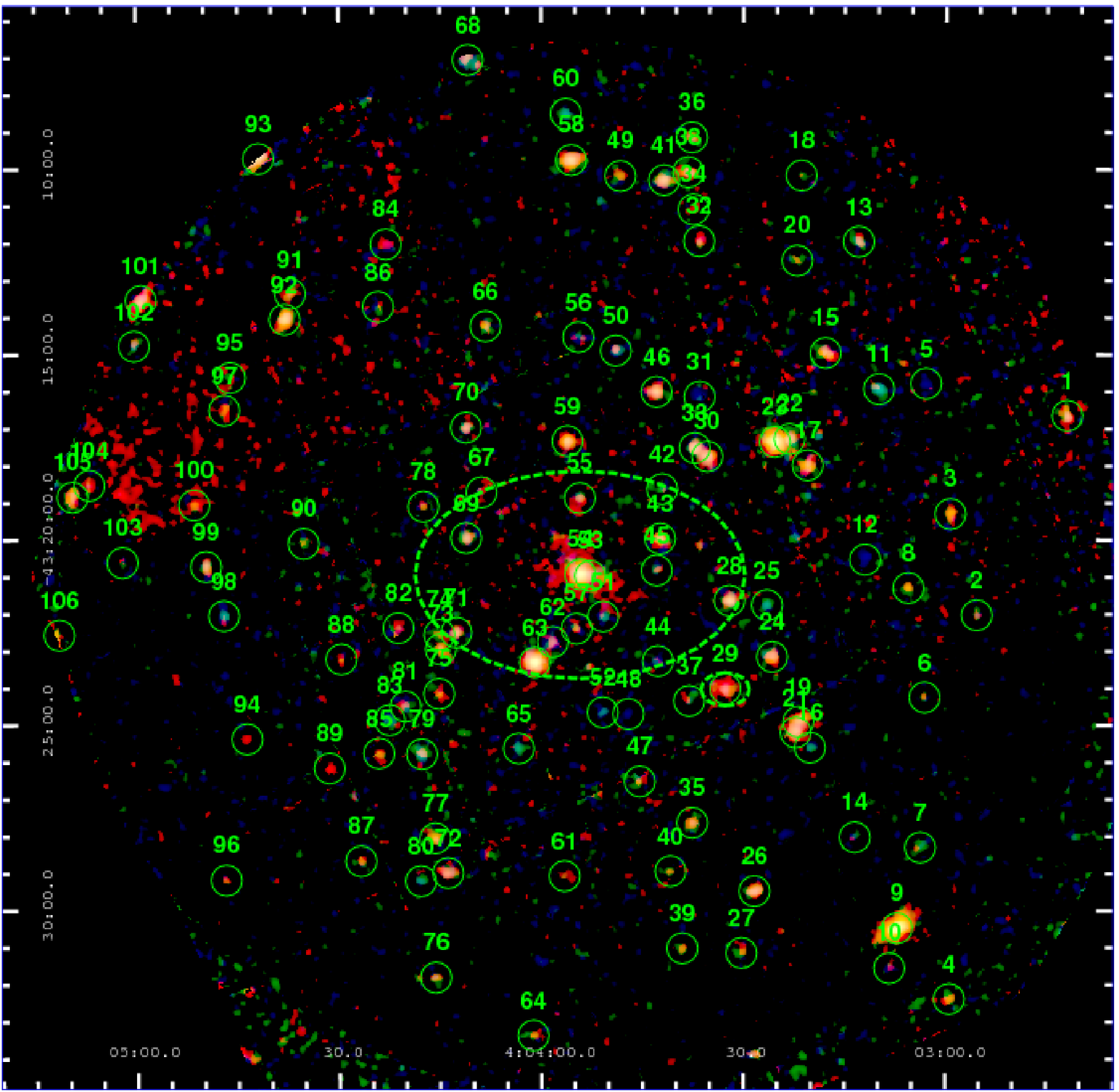}
\includegraphics[width=8cm]{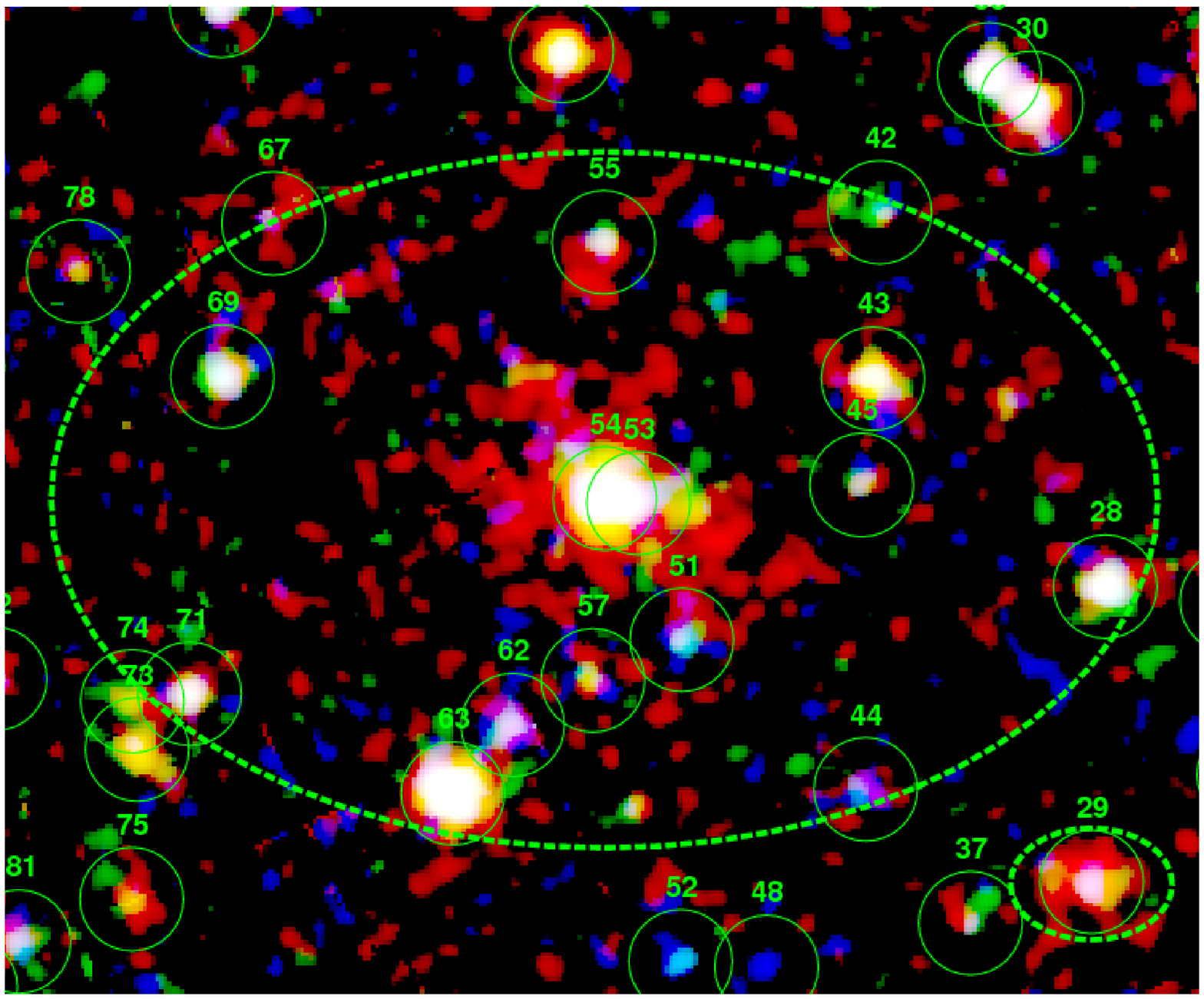}
\end{center}
\caption{Combined PN, MOS1, and MOS2 three-colour mosaic image of the NGC 1512/1510 field of view,
where red, green, and blue colors
represent the $0.2-1$~keV, $1-2$~keV, $2-4.5$~keV energy bands.
The green dashed ellipses are the $D_{25}$ ellipses for NGC~1512 and NGC~1510. 
The right panel shows a close-up view of the central region.}
\label{fig. xmm-mosaic}
\end{figure*}

\section{Observations and data reduction}
\label{sect. data analysis}

\subsection{\emph{XMM-Newton}: source detection}
\label{sect. data analysis xmm}

The galaxy pair NGC\,1512/1510 has been observed with \emph{XMM-Newton} (ID: 0693160101)
between 2012 June 16 (20:31 UTC) and 2012 June 17 (16:24 UTC)
in a single, 63\,ks exposure observation.
The data analysis was performed through the \emph{XMM-Newton}
Science Analysis System (SAS) software (version 12.0.1).
The observation was largely contaminated by high background due to proton flares.
After rejecting time intervals affected by high background,
the net good exposure time was reduced to $26.0$ ks for PN,
$39.8$ ks for Metal Oxide Semi-conductor 1 (MOS1), and $34.8$ ks for MOS2.

For each instrument, data were divided into five energy bands:
\begin{itemize}
\item $B_1$: 0.2--0.5 keV;
\item $B_2$: 0.5--1 keV;
\item $B_3$: 1--2 keV;
\item $B_4$: 2--4.5 keV;
\item $B_5$: 4.5--12 keV.
\end{itemize}
For the PN, data were filtered to include only single events
(PATTERN=0) in the energy band $B_1$,
and single and double events (PATTERN$\leq$4) for the other energy bands.
We excluded the energy range $7.2-9.2$~keV to reduce
the background produced by strong fluorescence lines in the outer detector area \citep{Freyberg04}.
For the MOS, single to quadruple events (PATTERN$\leq$12) were selected.

We ran the source detection using the SAS task {\tt edetect\_chain}
on the images corresponding to the five energy bands
and three instruments (total of 15 images) simultaneously.
The source detection procedure can be divided into three steps.
The first step creates temporary source lists and 
background maps for these source lists. Local background maps are then used
to detect the sources.
We adopted a minimum-detection likelihood of 7 to obtain the list of sources
(the detection likelihood $L$ is defined by $L= -\ln p$, where $p$ 
is the probability that a Poissonian fluctuation of the background
is detected as a spurious source).
To create the background maps the sources are removed from the images 
and a two-dimensional spline with 16 nodes
is fitted to the exposure-corrected image.
In the second step the background maps are used to improve the detection sensitivity
and hence to create a new source list. Here we adopted a minimum-detection likelihood of 4.
In the last step, the list of source positions obtained in the previous step
is used to perform a maximum-likelihood point-spread function (PSF) 
fit to the source count distribution simultaneously 
in all energy bands and each EPIC instrument (a description of this algorithm is given by 
\citealt{Cruddace88}). For this step we adopted a minimum likelihood of $L=6$.
This step provides the final source list.
The source detection procedure of the task {\tt edetect\_chain}
provides, for each detected source, many parameters such as 
the position, the positional error, 
count rate, likelihood of detection, and 
hardness ratios (see Table \ref{Tab. source list}).
We removed false detections (artefacts on the detectors
or diffuse emission structures) by visual inspection.
We detected 106 sources in the NGC 1512/1510 field of view.

Figure \ref{fig. xmm-mosaic} shows the combined PN, MOS1, and MOS2
three-colour \xmm\ image.
The numbers of the detected sources in Table \ref{Tab. source list}
are overplotted on the image.
The thick green ellipses are the $D_{25}$ ellipses 
for NGC~1512 and NGC~1510 \citep{deVaucouleurs91}
defined by the 25 mag arcsec$^{-2}$ B-band isophote.

\subsection{\emph{XMM-Newton}: diffuse emission}
\label{diff-imaging}
To search for extended X-ray emission in the \system\ system we used the 
\xmm\ Extended Source Analysis Software (XMM-ESAS), packaged in SAS 12.0.1. 
This program is based on the software used for the background modelling described 
in \citet{Snowden2004}. Essentially, XMM-ESAS consists of a set of tasks to 
produce images and spectra from observational data and to create quiescent 
particle background (QPB) images and spectra that can be subtracted from the 
observational science products \citep[see][]{Kuntz2008,Snowden2008}. 
To ensure compatibility of our analysis with the XMM-ESAS framework, 
we reprocessed the observational data according to the 
ESAS cookbook\footnote{Available at \burl{http://heasarc.gsfc.nasa.gov/docs/xmm/xmmhp\_xmmesas.html}}. 
Standard filtering and calibration were applied to the observational data using 
the XMM-ESAS tools \texttt{epchain}, \texttt{emchain}, \texttt{pn-filter}, 
and \texttt{mos-filter}. The CCDs of each of the EPIC instruments were then examined 
to ensure that none were operating in an anomalous state 
\citep[where the background at $E < 1$ keV is strongly enhanced, see][]{Kuntz2008}. 
We determined that CCD 4 of the EPIC-MOS1 was in an anomalous state and  
excluded it from further analysis. The tasks \texttt{pn-filter} and \texttt{mos-filter} 
clean the data of obvious soft-proton (SP) flares by calling the task \texttt{espfilt}. 
After filtering with \texttt{espfilt}, 16~ks of EPIC-PN data, and 27~ks and 25~ks of 
EPIC-MOS1 and EPIC-MOS2 data, respectively, were available for further analysis. 
Since this observation was highly affected by SP flares, the likelihood of substantial 
residual SP contamination is high. We estimated the level of this using the diagnostic 
tool of \citet{DeLuca2004}\footnote{Available at \burl{http://xmm2.esac.esa.int/external/xmm_sw_cal/background/epic\_scripts.shtml\#flare}}. 
Unfortunately, the filtered event lists of each of the EPIC instruments 
were flagged as being `extremely' contaminated by SPs. 
Therefore, we proceeded very tentatively and cautiously 
with our search for extended emission in the NGC~1512/1510 system.

We excluded point sources determined in Sect.~\ref{sect. data analysis xmm} from the observational data
using the SAS task \texttt{region} and the XMM-ESAS task \texttt{make\_mask}. 
The contour method of the  \texttt{region} task was used to define the exclusion regions. 
The task calculates the point spread function (PSF) at the source position and normalises 
for the source brightness. Source counts were then removed down to a PSF threshold of 
0.25 times the local background, that is, 
the point source is excluded down to a level where the surface brightness of the point source 
is one quarter of the surrounding background. This method has the advantage that the point source 
exclusion regions follow the brightness of the source, so that brighter sources have larger 
exclusion regions and shapes corresponding approximately to the source PSF. Hence, the number 
of diffuse counts is maximised by tailoring the point source exclusion regions to the individual sources. 
By excluding counts in this manner, contamination from the point sources is reduced to a negligible level.

\begin{figure*}
\begin{center}
\resizebox{\hsize}{!}{\includegraphics[bb=8 436 896 724, clip]{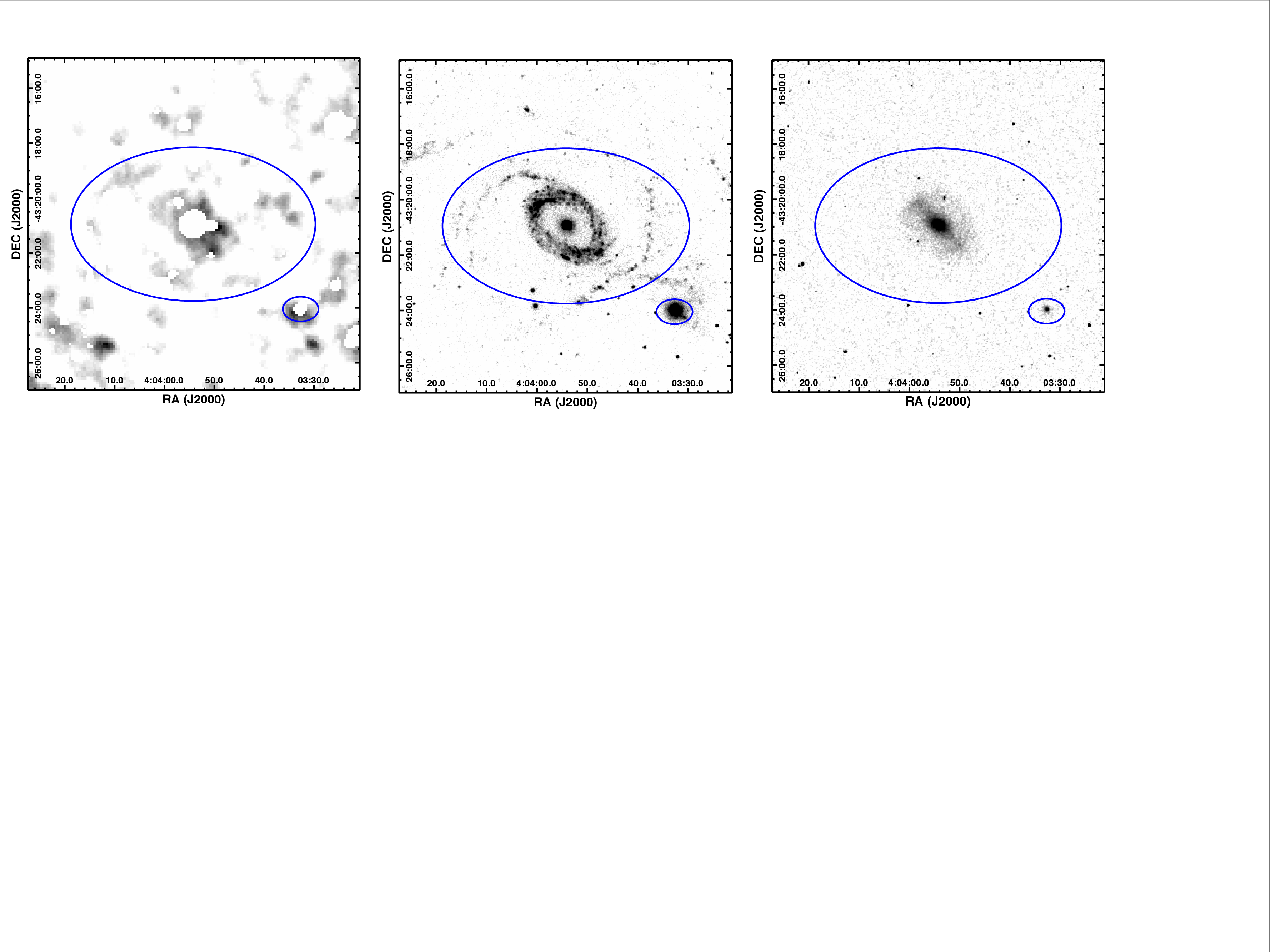}}
\caption{\textit{Left}: combined EPIC 0.3-2 keV image of the inner region of the NGC 1512/1510 observation. 
The image has been corrected for QPB and SP contamination, binned into 2x2 pixel bins before being adaptively 
smoothed using the XMM-ESAS task \texttt{adapt-900}. \textit{Middle}: \sw/UVOT image (uvm2 filter) 
of \system. \textit{Right}: 2MASS $K_{s}$-band image of \system\ taken from the 2MASS Large Galaxy Atlas 
\citep{Jarrett2003}. The blue ellipses represent the $D_{25}$ ellipses for the galaxy pair.  
}
\label{soft-inner-ngc1512}
\end{center}
\end{figure*}

The \texttt{pn-spectra} and \texttt{mos-spectra} were used to produce images with the point 
sources masked in the 0.3-2 keV and 2-7 keV energy bands. The \texttt{pn-back} and \texttt{mos-back} 
tasks were then used to produce corresponding QPB images. Using XMM-ESAS, it is also possible to model 
and retrospectively subtract residual SP contamination from the images. To this end we extracted 
full-field spectra and QPB backgrounds for the EPIC instruments and fitted the spectra in the 3$-$7 keV 
energy range where the SP contamination is expected to dominate (see Section \ref{pib} for a description 
of the SP fitting process). The determined spectral parameters were then used to generate model SP contamination 
images with the XMM-ESAS task \texttt{proton}. The \texttt{comb} task was used to merge the observational
QPB images and SP images from all three EPIC instruments into combined EPIC images. 
Finally, the \texttt{adapt-900} task was run to subtract the QPB and SP background, 
and adaptively smooth the resulting images. We note here that solar wind charge-exchange emission (SWCX), 
which is correlated to enhancements in the solar wind \citep{Snowden2004}, may also affect our observation. 
This is discussed in more detail in Section \ref{swcx}, but the effect is expected to be uniform across the FOV and, 
thus, it is only necessary to adjust the background flux levels. 
While no extended emission was observed in most of the \system\ field, there was evidence for faint soft emission 
in the innermost region of NGC~1512, illustrated in Figure~\ref{soft-inner-ngc1512}~(left). 
Morphologically, the diffuse X-ray emission appears to be associated with the region 
of the UV ring of \Gal\ \citep{Koribalski09}, shown in Figure~\ref{soft-inner-ngc1512}~(middle), 
which also corresponds to the region of higher stellar density, as evident from the $K_{s}$-band 
image taken from the 2MASS Large Galaxy Atlas\footnote{Available at \burl{http://irsa.ipac.caltech.edu/applications/2MASS/LGA/}} 
\citep{Jarrett2003}, Figure~\ref{soft-inner-ngc1512}~(right).

\subsection{Swift/XRT}
\label{sect. data analysis xrt}

NGC 1512 and  NGC 1510 have been observed with the
X-ray Telescope (XRT; \citealt{Burrows05}),
one of the three instruments on board
the \emph{Swift} satellite \citep{Gehrels04},
from July 2011 to January 2013, 
for a total exposure time of $\sim 33.6$ ks.

We used these data to study the long-term X-ray
variability of sources detected with both \emph{Swift}/XRT 
and \emph{XMM-Newton}/EPIC (see Sect. \ref{sect. variability}).
We collected the XRT observations in two groups:
\emph{obs1:} from 2011/07/26 00:13:25 to 2011/08/05 05:49:01;
\emph{obs2:} from 2012/09/25 01:09:59 to 2012/09/27 04:50:59 (see Table \ref{Tab. swift xrt log}).
We processed the XRT data obtained in photon-counting mode (PC)
with the standard procedures (xrtpipeline v0.12.6; \citealt{Burrows05}).
For each group of observations, source detection was performed
using \texttt{XIMAGE} (v4.5.1). For each source, we obtained the count rate and significance 
using the source statistics ({\tt sosta}) tool within \texttt{XIMAGE}.
The sources detected by \emph{Swift}/XRT with a signal-to-noise ratio $S/N>3$,
not located at the edge of the field of view (where the high background noise can be misinterpreted 
as real sources by the source detection procedure),
and with a unique \emph{XMM-Newton} counterpart are sources No. 54 
(the nuclear region of NGC 1512) and No. 63.
Owing to the low angular resolution of \emph{Swift}/XRT, 
the pairs of \emph{XMM-Newton} sources 30/33, 38/41, and 91/92 
have been detected by XRT as three sources.
Therefore, for the long-term variability study 
we only considered sources No. 54 and 63 (see Sect. \ref{sect. variability}).

\begin{table}
\begin{center}
\caption{\emph{Swift}/XRT observations of NGC 1512/1510.}
\label{Tab. swift xrt log}
\begin{tabular}{lcc}
\hline
\hline
ObsID       &       Start date    & $T_{\rm exp}$ (ks) \\
\hline
00045603001 & 2011-07-26 00:13:25 & 3077.8 \\
00045603002 & 2011-08-01 07:10:01 & 2880.3 \\
00045603003 & 2011-08-02 07:21:12 & 1965.2 \\	
00045603004 & 2011-08-04 00:59:00 &  767.2 \\
00045603005 & 2011-08-05 05:49:01 & 2447.1 \\
00045603008 & 2012-09-25 01:09:59 & 9497.0 \\
00045603009 & 2012-09-26 01:09:59 & 3778.5 \\  
00045603010 & 2012-09-27 04:50:59 & 4588.0 \\
\hline
\end{tabular}
\end{center}
\end{table}

\subsection{ATCA}
\label{sect. data analysis atca}

The system NGC~1512/1510 has been observed in radio (20-cm) with ATCA.
The observation consists of four pointings with multiple configurations
(see \citealt{Koribalski09} for the observing and data reduction details).

We used the 20-cm radio continuum maps with 8 and 15 arcsec resolution 
to find possible radio counterparts of the X-ray sources
we detected with \emph{XMM-Newton}.
We found nine radio counterparts of sources
13, 23, 29 (nuclear region of NGC\,1510), 54 (nuclear region of NGC\,1512), 67,
71, 83, 94, and 98. We used these associations to classify
five background objects (see Sect. \ref{sect. bg objects}).

\section{Analysis}
\label{sect. analysis}

\subsection{Astrometrical corrections}
\label{sect. astrometrical corrections}

We determined the systematic errors in the X-ray
positions of the \emph{XMM-Newton} observation
by calculating the offsets in the X-ray positions
of the \emph{XMM-Newton} sources identified and/or classified as
foreground stars and background objects (see Sect. \ref{sect. source classification}) 
with respect to their optical and infrared counterparts 
(USNO-B1 and 2MASS catalogues of 
\citealt{Monet03} and \citealt{Cutri03}).
In particular, we used the sources of Tables \ref{Tab. source-offset list}
and \ref{Tab. list-galaxies}, which were clearly identified as
foreground or background objects
except for source No.\,61, which appears to be extended on the DSS maps. 
The offset between the X-ray positions and the optical
positions corrected for proper motion is
$\Delta RA= -0.41 \pm 0.54$~arcsec, $\Delta Dec= -1.44 \pm 0.54$~arcsec.
Since the offset for $RA$ is not statistically significant,
we used only $\Delta Dec$ to correct the positions 
of all the detected sources.
The rms offsets in declination before and after correction are 2.4 and 2.1.

\subsection{Spectral analysis and hardness ratios of point sources}

\label{sect. spectral analysis and hardness-ratios}

We extracted the X-ray spectra of the sources
from circular or elliptical regions, which were adjusted by eye
for each source on each detector, depending on the presence of nearby sources.
Background spectra were extracted from source-free zones
and then normalised to the extraction area of the corresponding source.
The spectra, instrument responses, and ancillary files were generated
using the SAS software. We performed the spectral analyses 
using \texttt{XSPEC} (ver. 12.7.0, \citealt{Arnaud96}).

For brighter sources (counts $\gtrsim$ 300 in the energy range $0.2-12$ keV)
we used the $\chi^2$ statistics. The spectral energy channels were grouped to have
at least 20 counts per bin for good statistical quality of the spectra.

For sources with a fewer counts
we used the Cash statistic (or C-stat; \citealt{Cash79}), which allows
background subtraction in \texttt{XSPEC} by means of the 
W-statistic\footnote{see appendix 13 of \texttt{XSPEC} manual 
\url{http://heasarc.nasa.gov/xanadu/xspec/manual/XSappendixStatistics.html}}.
W-statistic requires at least one count per spectral bin.
We computed the quality of the fit with the \texttt{XSPEC} command
\texttt{goodness}, which performs Monte Carlo simulations of $10^4$ spectra
from the best-fit model and provides the percentage of simulated spectra
with a C-stat less than that obtained from the real data.
If the model represents the data accurately, the goodness should be close to 50\%.

We fitted the PN and MOS spectra simultaneously with different models.
A good fit with one of these models can be used
to classify the sources into one of the following classes:
\begin{itemize}
\item XRBs: power-law with $\Gamma=1-3$; disc-blackbody with $kT_{\rm in}=0.5-1$~keV (\citealt{White95}; \citealt{Makishima86});
\item supernova remnants: a thermal plasma model (e.g. \texttt{apec} of \citealt{Smith01} with $kT=0.2-1.5$~keV);
\item super-soft sources: blackbody with $kT_{\rm bb}=50-100$~eV \citep{DiStefano03}.
\end{itemize}

In total we fitted the spectra of eight sources.
With the exception of source No.~9 (see Sects. \ref{sect. fg stars} and \ref{sect. app. fgstars}),
the X-ray spectra can be fitted with an absorbed power-law
or an absorbed disc-blackbody model,
with photon indices and $kT_{\rm in}$ compatible with 
either XRBs or AGNs (see Sects. \ref{sect. bg objects} and \ref{sect. sources in NGC 1512/1510}).

For each source, we calculated four hardness ratios, defined as
\begin{equation} \label{eq. hr}
HR_i = \frac{B_{i+1} - B_i}{B_{i+1} + B_i} \mbox{ for } i=1,..., 4 .
\end{equation}
We calculated the hardness ratios of Eq. (\ref{eq. hr})
with the Bayesian method described in \citet{Park06}
and implemented in the \texttt{BEHR} code\footnote{See http://hea-www.harvard.edu/astrostat/behr/} (ver. 08-28-2012). 
This method is based on the assumption that the detection of X-ray photons 
is a random process described by Poisson statistics
instead of the classically assumed Gaussian statistic.
The hardness ratios calculated using the Bayesian method
are much more accurate than those calculated using the classical methods,
especially for very faint sources, for which the Poisson distribution becomes more asymmetric.
Moreover, if the source is not detected in one or more energy bands,
the Bayesian method of \citet{Park06} can produce reliable estimates
where classical methods would fail.
The required inputs for the Bayesian method of \citet{Park06} are
the source region counts, the background region counts
(both corrected for vignetting and out-of-time events), the ratio
of the source and background extraction areas, and the exposure time at the source position.

\begin{figure}
\begin{center}
\includegraphics[bb=75 371 560 825,clip,width=8cm]{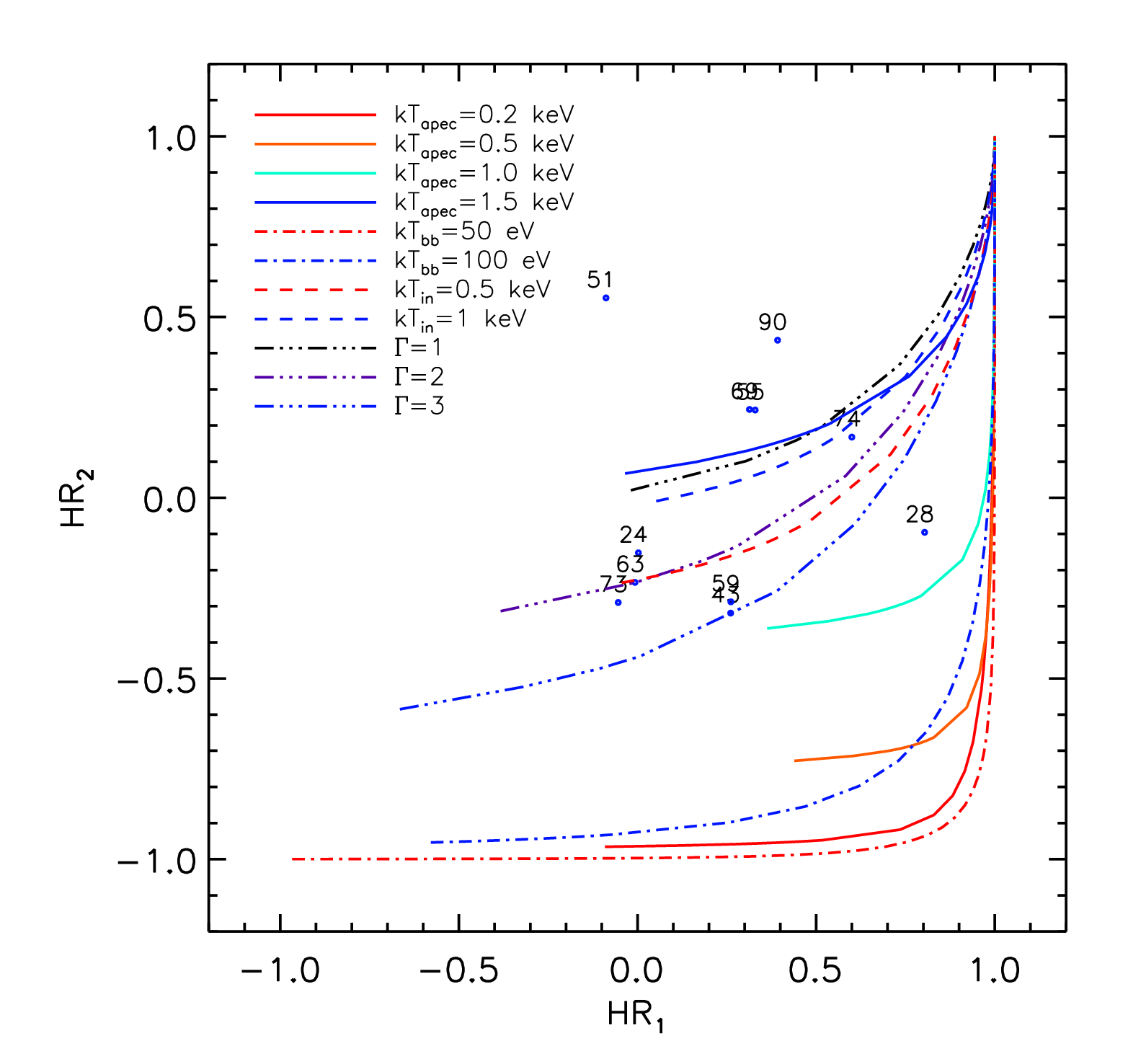}
\includegraphics[bb=75 371 560 825,clip,width=8cm]{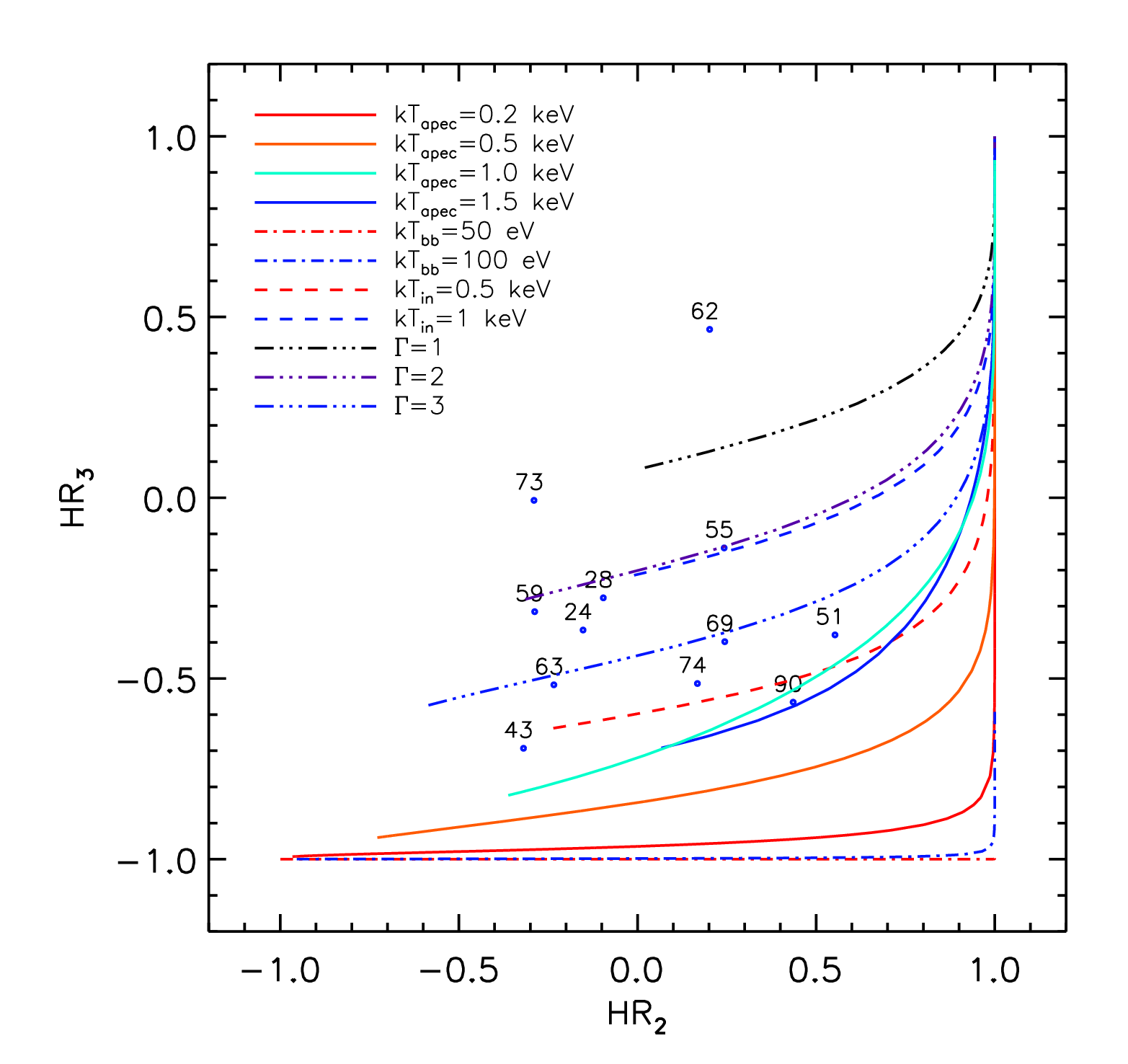}
\includegraphics[bb=75 371 560 825,clip,width=8cm]{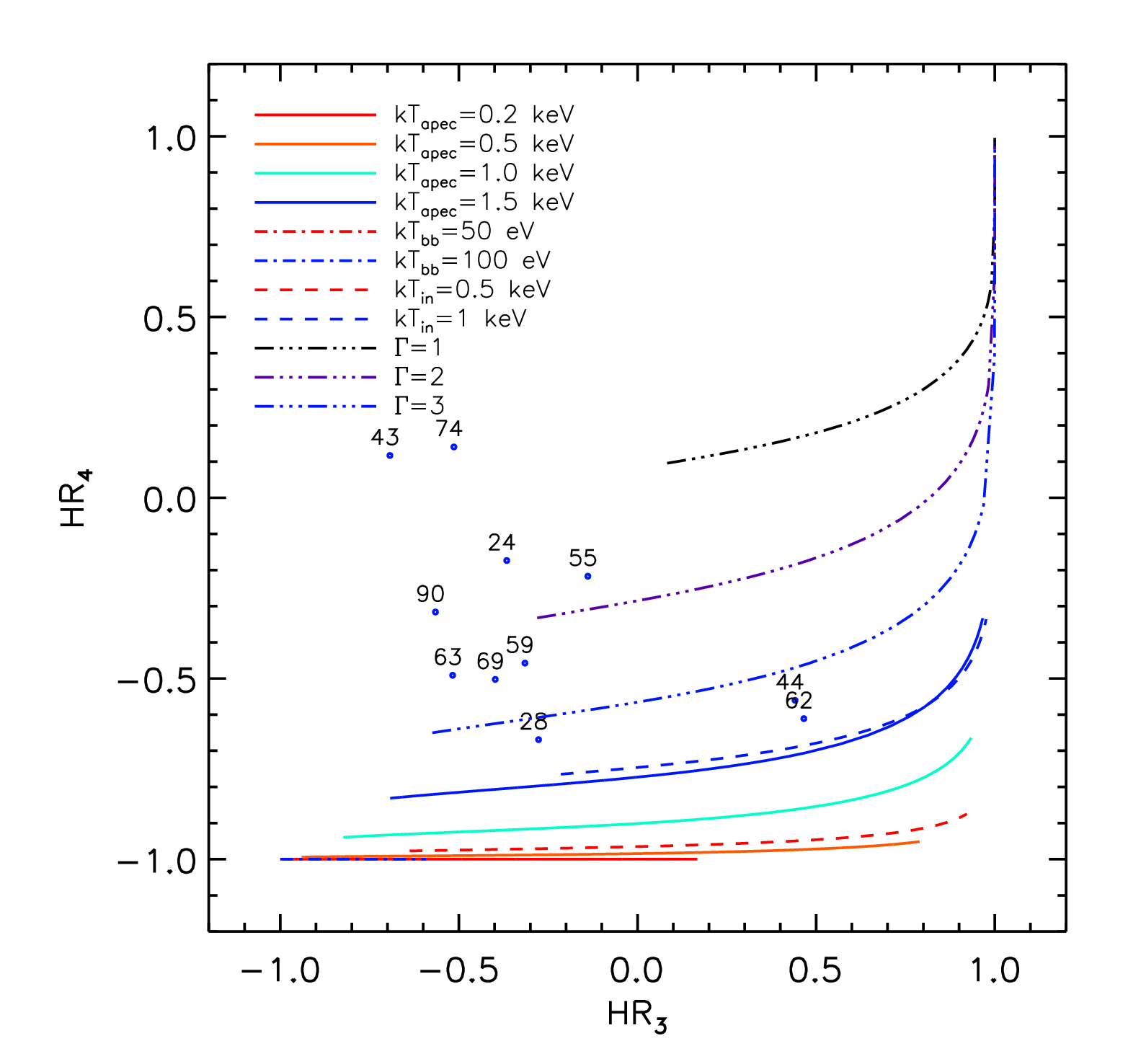}
\end{center}
\caption{Hardness-ratio diagrams of sources with error bars smaller than 0.3
detected in NGC 1512/1510.
The lines are the hardness ratios calculated for different spectral models and column densities,
as described in Sect. \ref{sect. spectral analysis and hardness-ratios}.
}
\label{fig. hr diagrams}
\end{figure}

Hardness ratios can provide information about the X-ray properties of faint sources, 
for which the spectral fitting is not possible (see e.g. \citealt{Prestwich03}).
The hardness ratios calculated for each source are listed in Table \ref{Tab. source list}.
Figure \ref{fig. hr diagrams} shows the hardness ratios 
of the sources detected by \emph{XMM-Newton} in NGC 1512/1510 
(see Sect. \ref{sect. sources in NGC 1512/1510}).
The grids of hardness ratios calculated for different spectral models,
with $N_{\rm H}$ ranging from $10^{20}$~cm$^{-2}$ to $10^{24}$~cm$^{-2}$,
can help to better separate different classes of sources.

\begin{table}
\begin{center}
\caption{Count rate to energy conversion factors for thin and medium filters of the
EPIC instruments in the energy ranges B1-B5, assuming an absorbed power-law
with a photon index of 1.7 and the Galactic foreground absorption $1.9 \times 10^{20}$ cm$^{-2}$
in the direction of NGC 1512/1510.}
\label{Tab. ecfs}
\resizebox{\columnwidth}{!}{
\begin{tabular}{l|cccccc}
\hline
\hline
Detector  & Filter &      B1     &      B2     &      B3    &       B4       &       B5    \\
EPIC      &        & \multicolumn{5}{c}{$(10^{-12})$ erg cm$^{-2}$ ct$^{-1}$}              \\
\hline
PN        & Medium &    1.070    &     1.106   &    1.725   &      4.969     &    17.47    \\
MOS       & Medium &    7.205    &     5.614   &    5.181   &     13.71      &    69.62    \\
\hline
\end{tabular}
}
\end{center}
\end{table}

\begin{table*}
\begin{center}
\caption{$0.5-10$ keV fluxes, variability factor and significance of the variability
of sources observed with \emph{XMM-Newton} and \emph{Swift}/XRT (flux in erg cm$^{-2}$ s$^{-1}$).}
\label{Tab. variability}
\begin{tabular}{lccccc}
\hline
\hline
\noalign{\smallskip}
No. &            obs1 (Jul-Aug 2011)        &                 Jun. 2012             &            obs2 (Sept. 2012)          & $V_{\rm f}$ &  $S$ \\
    &              (\emph{Swift})           &           (\emph{XMM-Newton})         &             (\emph{Swift})            &            &      \\
\noalign{\smallskip}
\hline
\noalign{\smallskip}
54  & $1.2{+1.9 \atop -0.6} \times 10^{-13}$ & $6.7{+2.0 \atop -2.0} \times 10^{-14}$ & $1.1{+0.6 \atop -0.5} \times 10^{-13}$ &     1.8    & 0.8 \\
\noalign{\smallskip}
63  & $7.5{+3.0 \atop -2.6} \times 10^{-14}$ & $7.4{+1.1 \atop -1.0} \times 10^{-14}$ & $1.2{+0.4 \atop -0.3} \times 10^{-13}$ &     1.6    & 1.4 \\
\noalign{\smallskip}
\hline
\end{tabular}
\end{center}
\end{table*}

\subsection{Variability analysis}
\label{sect. variability}

For each XMM-\emph{Newton} observation, 
we searched for pulsations of the brightest sources (counts\,$\gtrsim 100$)
on time scales between $\sim 0.15$ s and the time duration of each observation.
After extracting the event files, we applied both a Fourier transform and a $Z^2_n$ analysis
\citep{Buccheri83} for one to three harmonics. No statistically significant variability 
from the analysed sources was detected.

We studied the long-term X-ray variability of sources 
detected with both \emph{Swift}/XRT and \emph{XMM-Newton}/EPIC.
This was the case only for the two sources No. 54 and 63.
We calculated the \xmm\ fluxes with the energy conversion factors
reported in Table \ref{Tab. ecfs}.
Table \ref{Tab. variability} shows the $0.5-10$ keV \emph{Swift} and \emph{XMM-Newton} 
fluxes, variability factor $V_{\rm f}=F_{\rm max}/F_{\rm min}$
(where $F_{\rm max}$ and $F_{\rm min}$ are the highest and lowest fluxes)
and the significance of the variability $S=(F_{\rm max}-F_{\rm min})/\sqrt{\sigma^2_{\rm max} + \sigma^2_{\rm min}}$,
where $\sigma_{\rm max}$ and $\sigma_{\rm min}$ are the errors of the highest and lowest flux \citep{Primini93}. 
We derived the XRT fluxes 
from the count rates obtained with the tool \texttt{sosta} and 
the spectral parameters obtained in the \emph{XMM-Newton} 
data analysis, assuming no changes in the spectral shape 
(see Table \ref{Tab. no.23 54 63 spectral parameters}).
Sources No. 54 and 63 do not show a significant ($S \geq 3$) variability.

\subsection{Spectral analysis of diffuse emission}
\label{sect. sp. analysis diff. emission}

The spectral analysis of the region of soft diffuse emission in the centre of \Gal\ is complicated 
because of the high residual SP contamination and the possible contribution of fainter, 
large-scale emission in \Gal\ and/or SWCX emission. While these may not be so problematic for bright 
extended emission, the faint nature of the central diffuse emission required us to treat the background 
with utmost care to extract the purest possible spectral results. Since any additional faint 
contribution from \Gal\ most likely varies across the FOV, a straightforward background extraction and 
subtraction from a nearby region may not be appropriate. In addition, backgrounds selected in this 
manner from different regions of the detector with different responses may also compromise our results. 
For these reasons, we decided to use XMM-ESAS techniques for the spectral analysis, that is, we subtracted 
the modelled QPB spectra and accounted for the remaining X-ray background 
by including a physically motivated model in the fits.

The XMM-ESAS tasks \texttt{pn-spectra} and  \texttt{mos-spectra} were used to extract the spectra 
and response files from the emission region within the ring of \Gal\ (see Fig. \ref{soft-inner-ngc1512}). 
\texttt{pn-back} and \texttt{mos-back} were used to produce corresponding QPB spectra to be subtracted 
from the observational spectra. The spectra, associated response files, and QPB spectra were linked 
using the \texttt{grppha} task in the FTOOLS package\footnote{\burl{http://heasarc.gsfc.nasa.gov/ftools/}}. 
The spectra were grouped to at least 30 counts per bin to allow the use of the $\chi^{2}$-statistic. 
All fits were performed using \texttt{XSPEC} \citep{Arnaud1996} version 12.7.1 with ATOMDB\footnote{http://www.atomdb.org/} 
version 2.0.1, abundance tables set to those of \citet{Wilms2000}, and photoelectric absorption cross-sections 
set to those of \citet{Bal1992}. We limited our analysis to $>0.4$~keV to avoid the to strong low-energy tail 
due to electronic noise, as recommended in the ESAS cookbook. Following the subtraction of the QBP spectrum, 
we accounted for the remaining constituents of the X-ray background in \texttt{XSPEC}. 
These fall into two categories: the particle-induced background and the astrophysical background.

\subsubsection{Particle-induced background}
\label{pib}
After subtracting the QPB, the remaining particle-induced background consists of instrumental 
fluorescence lines and the residual SP contamination. The instrumental fluorescence lines were
modelled with Gaussian components ({\tt gauss} in \texttt{XSPEC}) at 1.49 keV in the EPIC-PN spectrum, and 1.49 keV 
and 1.75 keV for the EPIC-MOS spectra. Additional instrumental lines are present at harder energies 
in the EPIC spectra, but because we limited our analysis to the 0.4-5 keV energy range, these were ignored. 
The residual SP contamination was modelled with a power-law ({\tt powerlaw} in \texttt{XSPEC}) component not folded 
through the instrumental response. This is possible by using the diagonal response matrices supplied in the 
CALDB of XMM-ESAS. Given the small extraction area of the inner \Gal\ region, the low number of counts at 
harder energies prohibit robust constraints on the SP contamination model, most importantly on the slope 
of the power-law component. This is a problem because uncertainties on this parameter 
will dramatically affect the derived diffuse emission spectral parameters. 
To circumvent this we used the full-field EPIC spectra (see Sect. \ref{diff-imaging}), 
excluding point sources, and fitted these spectra in the 3--7~keV energy range. Because the SP contamination 
component should dominate at these energies, we can constrain the shape of the spectrum, which was then fixed 
in the fits to the inner \Gal\ spectrum. We find that the residual SP contamination can be modelled as a power-law 
of photon index $\Gamma=0.22$ for the EPIC-PN spectrum and $\Gamma=0.35$ for the EPIC-MOS spectra.

\subsubsection{Astrophysical background}
The astrophysical background (APB) typically comprises four or fewer components \citep{Snowden2008,Kuntz2010}, 
namely the unabsorbed thermal emission from the local hot bubble (LHB, $kT\sim$ 0.1 keV), 
absorbed cool ($kT\sim$ 0.1 keV) and warm ($kT\sim$ 0.25 keV) thermal emission from the Galactic halo, 
and an absorbed power-law representing unresolved background cosmological sources \citep[$\Gamma \sim 1.46$,][]{Chen1997}. 
All thermal components were fitted with the {\tt vapec} \citep{Smith01} thermal plasma model in \texttt{XSPEC}. 
To model the absorption of the Galactic halo and cosmological background sources we used the photoelectric 
absorption model {\tt phabs} in \texttt{XSPEC}. The value of the foreground hydrogen absorption column 
was fixed at $1\times10^{20}$ cm$^{-2}$ based on the Leiden/Argentine/Bonn (LAB) Survey of Galactic HI \citep{Kalberla2005}, 
determined using the HEASARC $N_{\rm{H}}$ Tool\footnote{\burl{http://heasarc.gsfc.nasa.gov/cgi-bin/Tools/w3nh/w3nh.pl}}. 
Because of this very low foreground absorption value, the LHB and cool Galactic halo emission are more or less 
indistinguishable in our spectral analysis energy range, which is why we treated them as a single component in the fits.

\subsubsection{SWCX}
\label{swcx}
\par
\begin{figure*}
\begin{center}
\resizebox{\hsize}{!}{\includegraphics[bb=7 384 952 733, clip=true, angle=0]{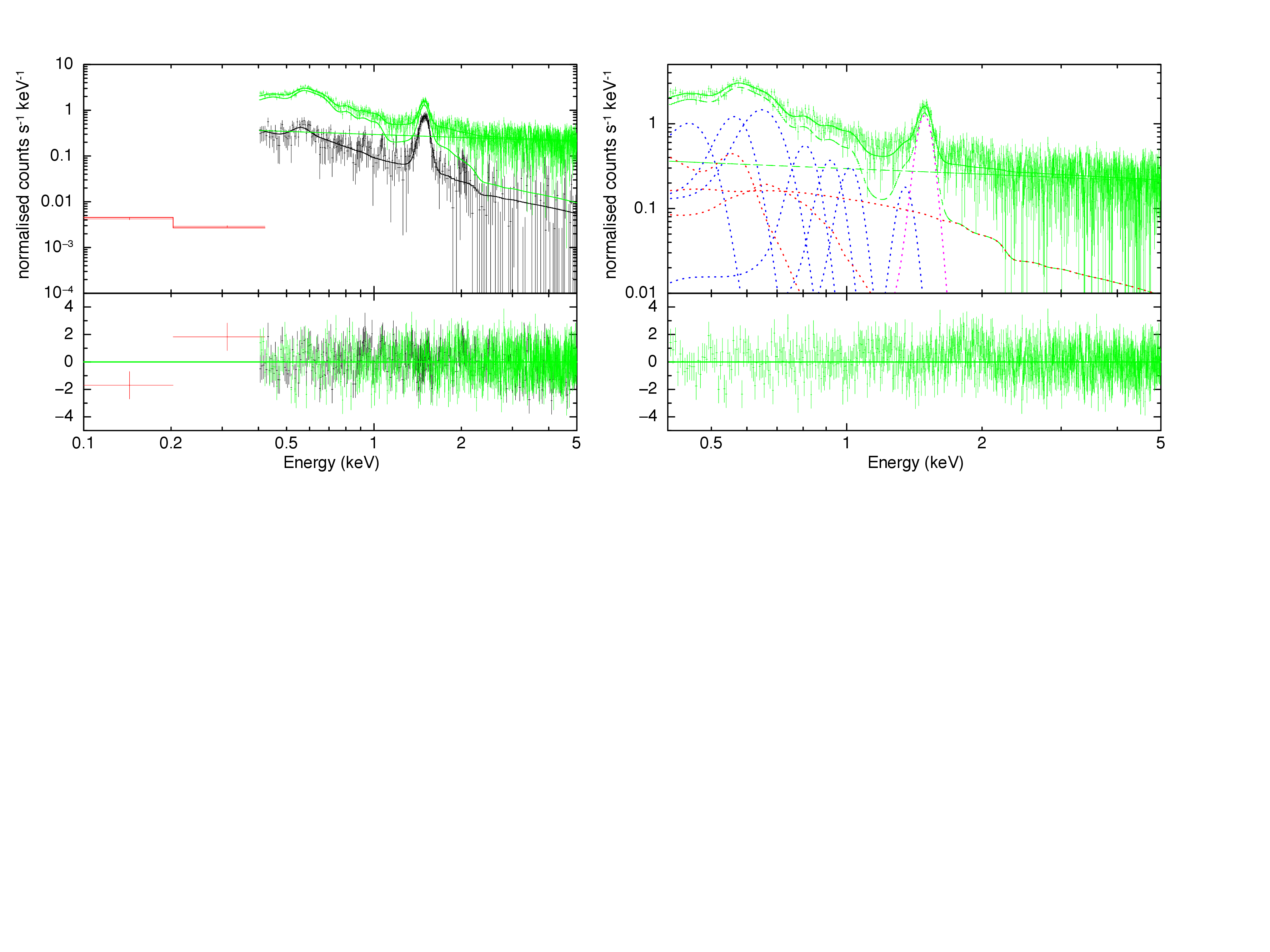}}
\caption{\textit{Left}: Various background spectra considered for our analysis. The black lines indicate the background field 
spectrum that was fitted simultaneously with the RASS background (shown in red) to constrain the emission components $<0.5~$keV. 
This model was then used to fit the \system\ observational background (shown in green for comparison), which required additional 
components for SWCX contamination. \textit{Right}: \system\ observational background. Additional model components are indicated with 
APB in red, fluorescence line in magenta, and SWCX lines in blue. The green dashed line that follows the data more closely represents 
the combined additive model, while the straight dashed black line represents the residual SP contamination. The addition of these 
two provides the total model indicated by the solid green line. In each panel, only EPIC-PN spectra are shown for clarity.
}
\label{bg-spectra}
\end{center}
\end{figure*}

 Given the high level of residual SP contamination in our observation, we also considered whether 
SWCX emission is present in our spectra as well. An SWCX spectrum consists of emission lines corresponding to 
the ions in the solar wind. Unfortunately, these lines are also of great astrophysical importance and can 
affect the derived spectral parameters of the astrophysical source at issue. SWCX occurs in several regions 
of the solar system, but the most important in terms of \xmm\ observations is caused by solar wind ions
that interact with the Earth's magnetosheath \citep{Robertson2003a}. There are two main influences on the 
level of SWCX contamination of a spectrum. First, because SWCX is generated by highly charged ions in the solar wind, 
flux enhancements of the solar wind naturally lead to a higher production of SWCX X-rays. Second, 
the viewing geometry of \xmm\ with respect to the magnetosheath will affect the level of contamination, 
meaning that observations made with \xmm\ through more of the magetosheath are subject to more contamination. 
Because of the orbit of \xmm\ around Earth and that of Earth around the Sun, certain periods of the year are more 
susceptible to SWCX contamination, with the summer months being the worst affected 
\citep[see Fig. 1 of ][for a nice illustration]{Carter2008}. Our observation of \Gal\ with strong residual 
SP contamination was performed on June 16/17 2012 and therefor might be affected. The easiest way to identify 
SWCX contamination is to have multi-epoch observations of the same region of sky. We searched the \xmm\ archives 
for an appropriate observation. Obs ID. 0501210701 (PI: M Ajello) was performed on July 7 2007, 
has an aimpoint $\sim30\arcmin$ from the nucleus of \Gal, and was found to be effectively free of SP contamination. 
Using XMM-ESAS, we extracted spectra from the central $10\arcmin$ of this `background' field and from all of the 
\system\ field (point sources and central diffuse emission region excluded) to characterise the APB and search for SWCX. 
Because we limited our \xmm\ spectral analysis to $>0.4$~keV, we obtained the $ROSAT$ All-Sky Survey (RASS) 
spectrum from a 1$^{\circ}$-2$^{\circ}$ annulus around \Gal\ (using the HEASARC X-ray background 
tool\footnote{\burl{http://heasarc.gsfc.nasa.gov/cgi-bin/Tools/xraybg/xraybg.pl}}) to constrain the soft emission 
below 0.5~keV. We initially fitted the background field spectra with the RASS spectrum to constrain the local APB 
(see the black and red spectra in Fig. \ref{bg-spectra}-left). We obtained a good fit to the data ($\chi^{2}_\nu$=1.05) 
with fit parameters in the expected ranges. Next we assumed that this background model represented the background of the 
\system\ observation and fixed the relevant background components in the model. The resulting fit to the spectra 
was very poor ($\chi^{2}_\nu>$2) with high residuals $<1$~keV, which could be attributed to SWCX. Therefore, 
we added a series of Gaussian components to the model that represent the emission lines expected from SWCX 
(listed in the ESAS cookbook), namely  C~VI (0.46~keV), O~VII (0.56~keV), O~VIII (0.65~keV), O~VIII (0.81~keV), 
Ne~IX (0.92~keV), Ne~IX (1.02~keV), and Mg~XI (1.35~keV). The addition of these emission lines resulted in a substantially 
improved fit ($\chi^{2}_\nu$=1.11), which is shown in green in Fig.~\ref{bg-spectra}~(left) for comparison with 
the background field and RASS spectra, and alone in Fig.~\ref{bg-spectra}~(right) with the additive components of 
the spectral model highlighted. Based on these results, it is evident that the observation of \system\ is heavily 
contaminated by SWCX. While there may be some real \Gal\ emission buried in this contamination, the level of 
SWCX prohibits any meaningful treatment. Accordingly, we considered only the central diffuse emission region for  
the subsequent analysis with the background components (APB, SWCX, and SPs) fixed appropriately.

\subsubsection{Spectral fitting}
We extracted EPIC-MOS and EPIC-PN spectra from the central region of diffuse emission in \Gal. The count statistics for each spectrum 
are relatively low, with EPIC-MOS spectra $\sim$~400 counts and EPIC-PN $\sim$~650 counts. 
Hence, we decided to fit only the EPIC-PN data 
because this spectrum is of the best statistical quality. 
While not completely satisfactory, it at least allowed us to obtain coarse 
estimates of the diffuse emission parameters, although more complex models 
(e.g., non-equilibrium ionisation components, metallicity enhancements) 
are beyond our grasp. We assumed that the soft emission in the central \Gal\ region is thermal in origin. 
Therefore, we fitted the EPIC-PN spectrum with the fixed background components 
and a single thermal plasma model ({\tt vapec} in \texttt{XSPEC}) 
with abundances set to those of \Gal\ \citep[0.65 solar,][]{Koribalski09}.  
Absorption components were included to represent the foreground 
Galactic column density and absorption due to material in \Gal. 
This second component consistently tended to zero and only an upper limit could be determined. 
However, we caution that the fit is not very sensitive to this parameter since the SWCX lines dominate 
the softer energies where the constraints on this parameter are determined. 
Thus, the resulting upper limit may be somewhat misleading.
Comparing this upper limit with the determined $N_{\rm{H}}$ values of the compact sources in \Gal\ 
(see Table \ref{Tab. no.23 54 63 spectral parameters}), we see that the upper limit from the diffuse emission 
fit is an order of magnitude higher than that of all sources in the $D_{25}$ ellipse of \Gal. 
The $N_{\rm{H}}$  values of the compact sources seem to point to a very low intrinsic absorption in the galaxy. 
The resulting spectral fit was statistically 
acceptable ($\chi^{2}_\nu$=0.93) with $kT=0.66~(0.28-0.89)$~keV and unabsorbed 
flux of $1.1 \times 10^{-14}$~erg~cm$^{-2}$~s$^{-1}$. Because this flux is derived 
from diffuse emission regions with point sources masked, we corrected it for the excluded regions. 
Assuming that the average surface brightness of the detected diffuse emission 
holds across the point source regions, the resulting X-ray flux 
is $F_{\rm{X,0.3-10~keV}} = 1.4 \times 10^{-14}$~erg~cm$^{-2}$~s$^{-1}$. 
This corresponds to an X-ray luminosity $L_{\rm{X,0.3-10~keV}} = 1.5 \times 10^{38}$~erg~s$^{-1}$ at the distance of \Gal.

\begin{table}[htdp]
\caption{Spectral fit results for the central diffuse emission in \Gal. See text for description of the model components.}
\begin{center}
\label{fit-results}
\resizebox{\columnwidth}{!}{
\begin{tabular}{lll}
\hline
\hline
Comp. & Parameter  & Value  \\
\hline
\multicolumn{3}{c}{Foreground Absorption}\\
\hline
\texttt{phabs} & $N_{\rm{H}}$ ($10^{22}$ cm$^{-2}$) & 0.01 (fixed)\tablefootmark{a} \\
\texttt{vphabs}& $N_{\rm{H}}$ ($10^{22}$ cm$^{-2}$) & 0.00 ($<0.38$))\tablefootmark{b} \\
& & \\
\hline
\multicolumn{3}{c}{\Gal\ components}\\
\hline
\texttt{vapec}\tablefootmark{c} & $kT$ (keV) & 0.66 (0.28--0.89) \\
& norm ($10^{-6}$) & 1.28 (0.84--1.76) \\
& $F_{\rm{X,0.3-10~keV}}$ ($10^{-14}$ erg~cm$^{-2}$~s$^{-1}$) & 1.4\tablefootmark{d} \\
& $L_{\rm{X,0.3-10~keV}}$ ($10^{38}$ erg~s$^{-1}$) & 1.5\tablefootmark{d} \\
& & \\
\hline
Fit statistic & $\chi^{2}_\nu$ & 0.93 (27 d.o.f.)\\
& & \\
\hline
\end{tabular}
}
\tablefoot{
The APB, SWCX, and SP parameters are fixed according to the background results described in the text. \\
\tablefoottext{a}{Fixed to the Galactic column density from the Leiden/Argentine/Bonn (LAB) Survey of Galactic HI \citep{Kalberla2005}.}
\tablefoottext{b}{Abundances fixed to those of \Gal. Only an upper limit for $N_{\rm H}$ could be determined. We caution that this upper limit may be misleading as the fit is not very sensitive to the parameter (see text).}
\tablefoottext{c}{Abundances fixed to those of \Gal.}
\tablefoottext{d}{Values are de-absorbed.}
}
\end{center}
\end{table}%

\begin{figure}
\begin{center}
\resizebox{\hsize}{!}{\includegraphics[bb= 17 407 473 744, clip=true, angle=0]{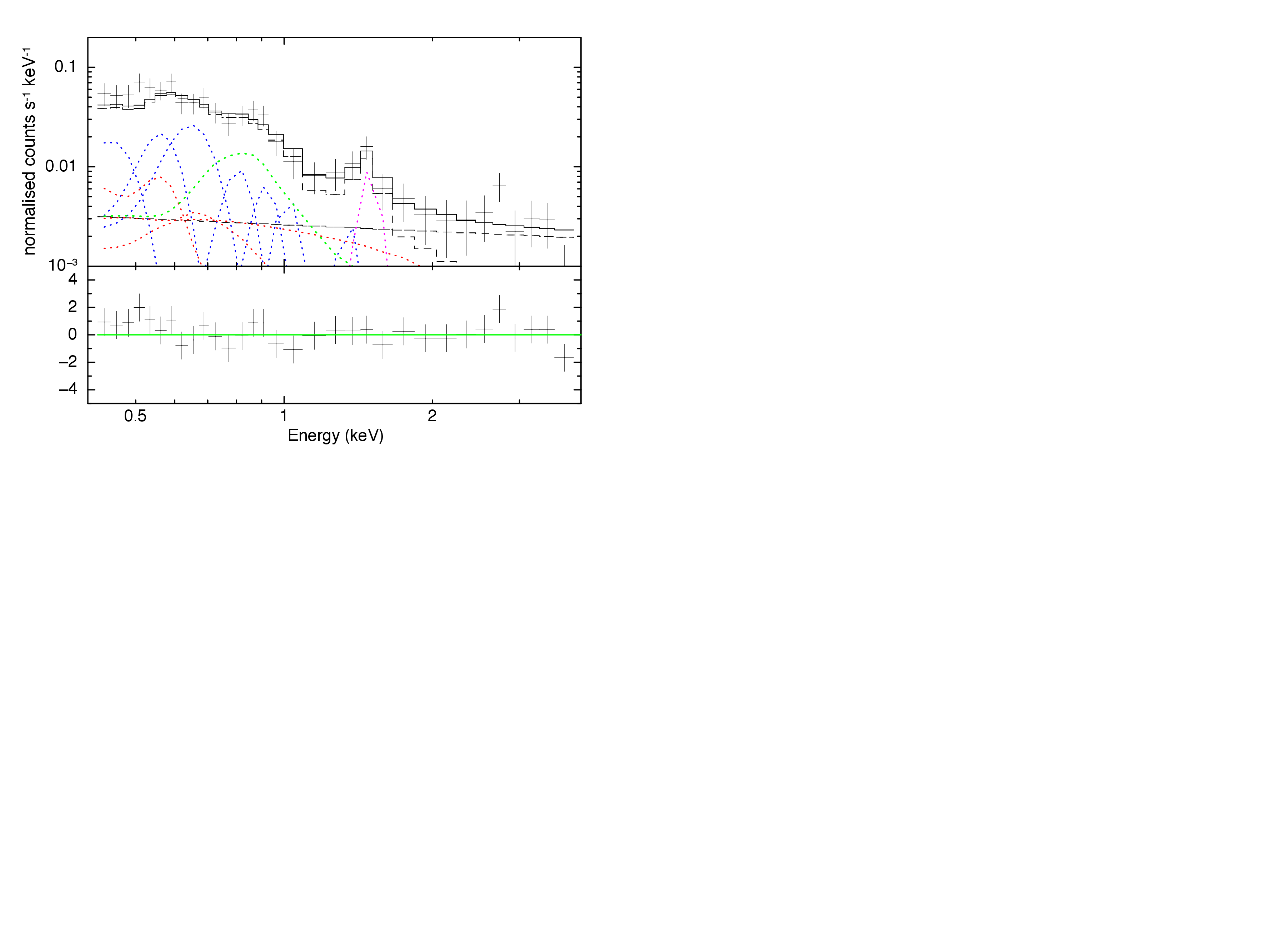}}
\caption{Fit to the EPIC-PN spectrum of the central diffuse emission in \Gal. Additive model components are indicated with APB in red, fluorescence line in magenta, SWCX lines in blue, and source emission in green. Results of the source component are presented in Table~\ref{fit-results}. The black dashed line that follows the data more closely represents the combined additive model, while the straight dashed black line represents the residual SP contamination. The addition of these two provides the total model indicated by the solid black line.
}
\label{pn-fit}
\end{center}
\end{figure}

\section{Source classification}
\label{sect. source classification}

We compared the list of X-ray sources detected in the \emph{XMM-Newton} observation
with sources observed in other wavelengths to find the counterparts.
To this purpose we used optical (USNO-B1, \citealt{Monet03}; 
Muenster Red Sky Survey (MRSS), \citealt{Ungruhe03}; Tycho-2 Catalogue (TYC), \citealt{Hog00}), 
infrared (2MASS, \citealt{Skrutskie06}), 
and radio (Sydney University Molonglo Sky Survey (SUMSS), \citealt{Mauch03}) catalogues.
We considered sources observed at different wavelengths to be associated
if their positions were closer than the three times combined statistical errors.
When an X-ray source had many optical counterparts,
we only considered the brightest one within the error circle.
This comparison allowed us to identify some of the \xmm\ sources
with sources previously classified in other wavelengths.
When a previous classification was not available, 
we tried to classify the X-ray sources
using their X-ray, optical, infrared, and radio properties.

For each X-ray source, possible optical, infrared, and radio counterparts are 
reported in Table \ref{Tab. source list classification}.

\subsection{Foreground stars}
\label{sect. fg stars}

\begin{figure*}
\begin{center}
\includegraphics[bb=92 360 564 841, clip, width=9.1cm]{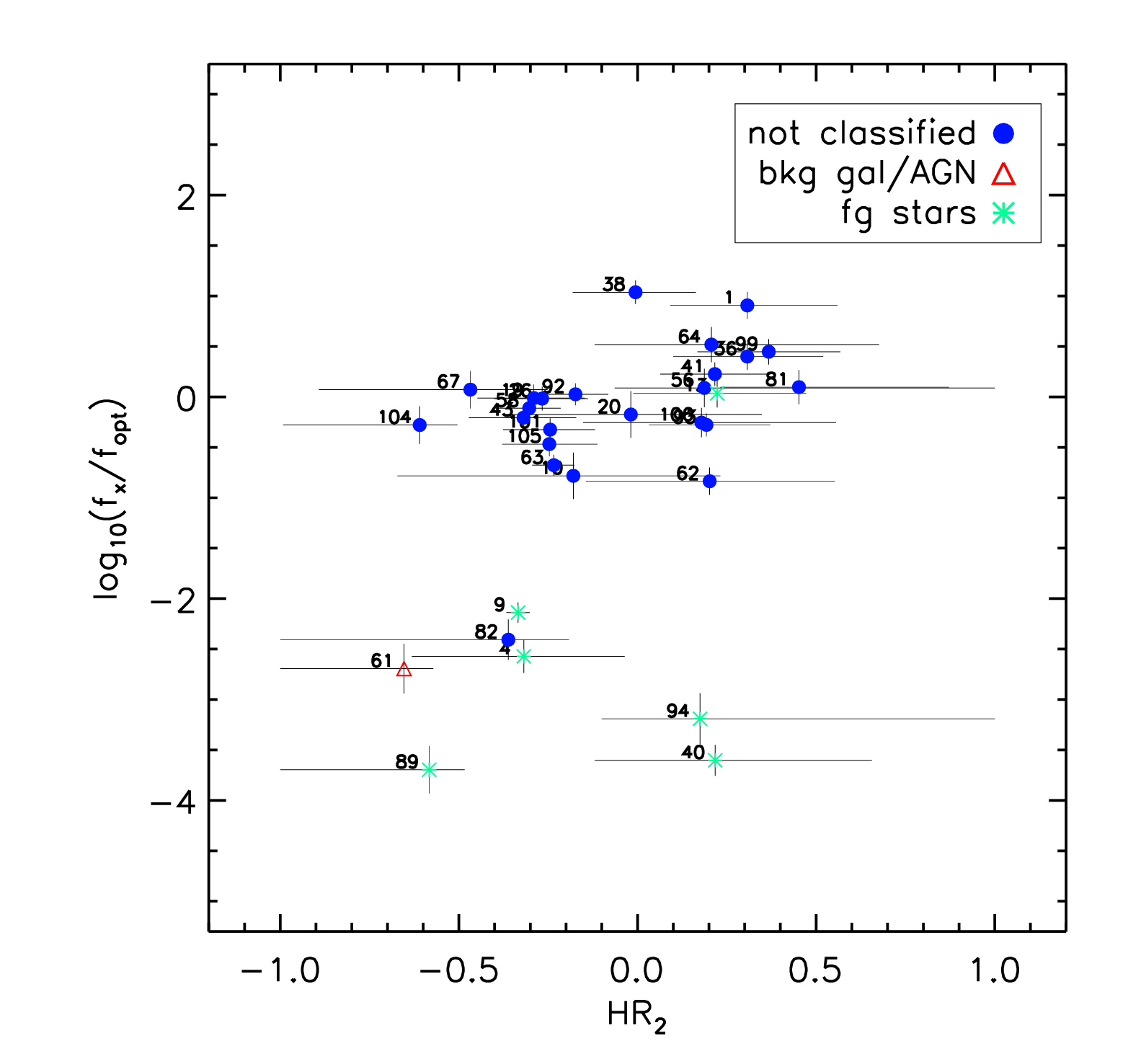}
\includegraphics[bb=92 360 564 841, clip, width=9.1cm]{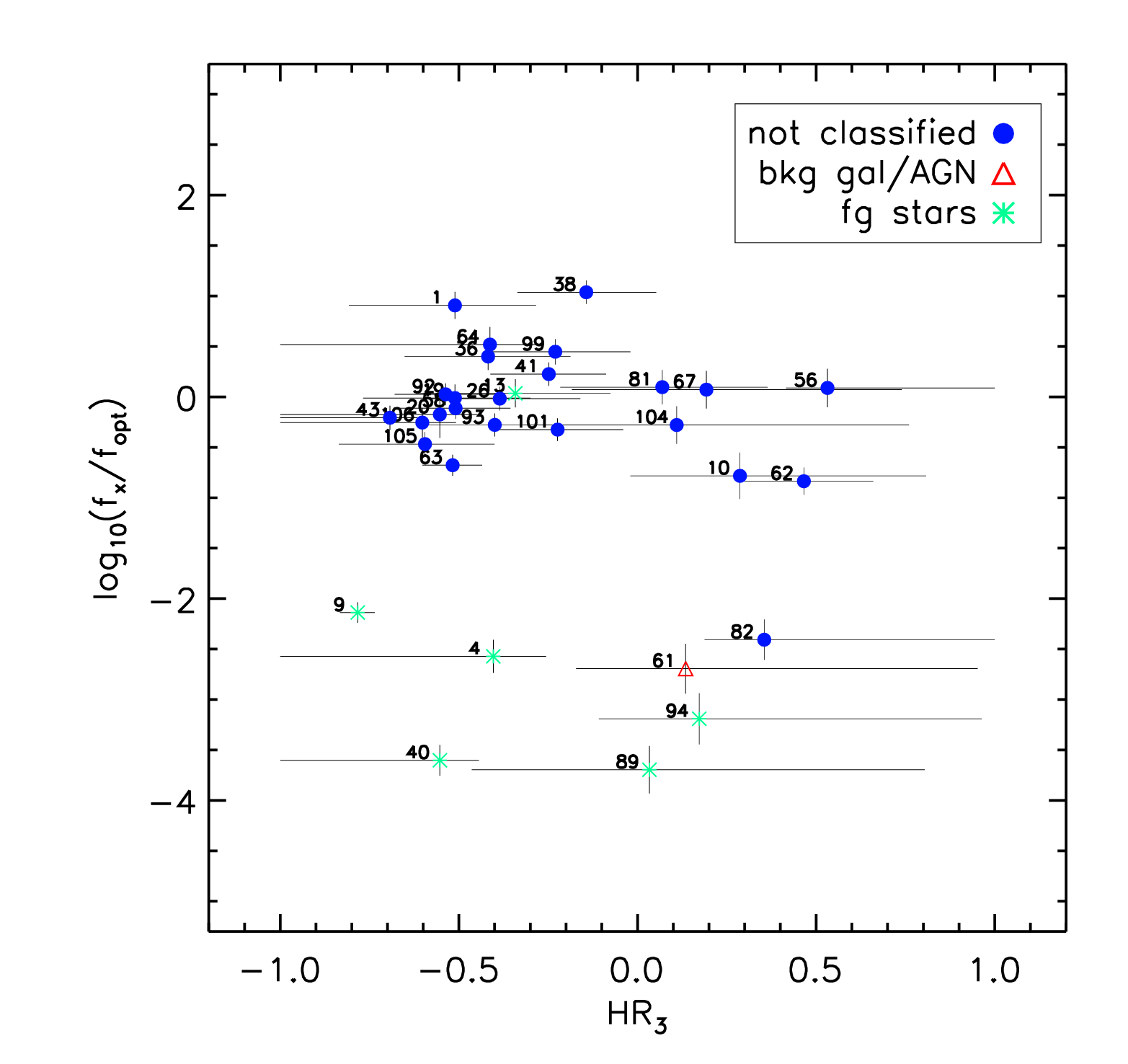}
\end{center}
\caption{Flux ratio $\log (f_{\rm x}/f_{\rm opt})$ over hardness ratios HR$_2$ and HR$_3$.}
\label{fig. log10-fx-fopt}
\end{figure*}

\begin{figure}
\begin{center}
\includegraphics[bb= 92 360 564 841, clip, width=9cm]{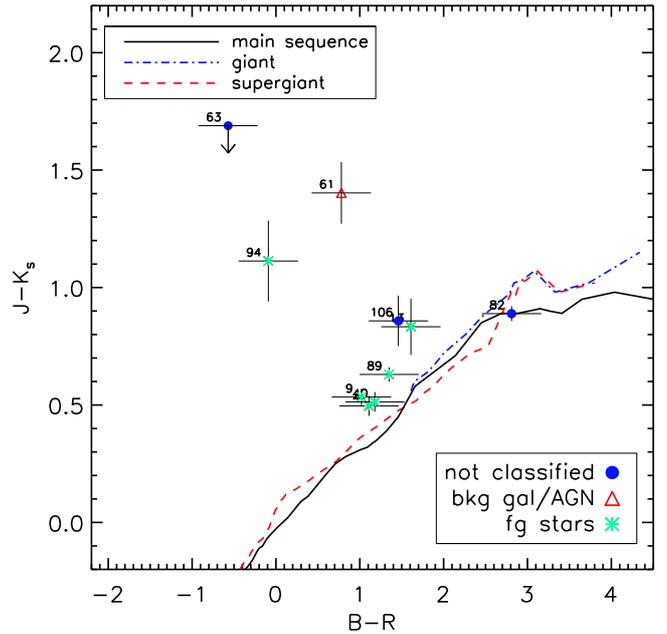}
\end{center}
\caption{Colour-colour diagram of \emph{XMM-Newton} sources with optical (USNO-B1) and infrared (2MASS) counterparts.}
\label{fig. b-r_j-k.ps}
\end{figure}

The \xmm\ field of view of NGC~1512/1510 is contaminated by stars
belonging to our Galaxy. The X-ray spectra of foreground stars
are somewhat softer than the X-ray spectra of other classes of sources.
Moreover, their X-ray emission is usually expected to be 
$f_{\rm x} \lesssim 10^{-1} f_{\rm opt}$ (e.g. \citealt{Krautter99}),
where $f_{\rm opt}$ is the flux in the optical band.
A useful tool to classify stars
compares the observed
X-ray-to-optical and X-ray-to-infrared flux ratios with 
the expected ones.
For each X-ray source with optical or infrared counterparts,
we calculated $f_{\rm x}/f_{\rm opt}$ and $f_{\rm x}/f_{\rm IR}$
with the equations of \citet{Maccacaro88}
and \citet{Lin12}:
\begin{eqnarray}
\log_{10}(f_{\rm x}/f_{\rm opt}) & = & \log_{10}(f_{\rm x}) + \frac{m_{\rm v}}{2.5} +5.37 \label{eq. maccacaro}\\
\log_{10}(f_{\rm x}/f_{\rm IR}) & = & \log_{10}(f_{\rm x}) + \frac{m_{\rm K_s}}{2.5} +6.95 \mbox{\ ,} \label{eq. lin}
\end{eqnarray}
where $m_{\rm v}$ is the visual magnitude
that we obtained from the USNO-B1
red and blue magnitudes, assuming $m_v \approx (m_{\rm red} + m_{\rm blue})/2$
and $m_{\rm k_s}$ is the magnitude in the $K_{\rm s}-$band from the 2MASS catalogue.
Figure \ref{fig. log10-fx-fopt} shows $\log_{10}(f_{\rm x}/f_{\rm opt})$ over the hardness ratios 
$HR_2$ and $HR_3$.
Foreground stars are expected to be in the regions of Fig. \ref{fig. log10-fx-fopt} 
with $\log_{10}(f_{\rm x}/f_{\rm opt}) \leq -1$ and, because of their soft X-ray spectra, 
$HR_2 \lesssim 0.3$ and $HR_3 \lesssim -0.4$ \citep{Pietsch04}.
Following the classification scheme derived by \citet{Lin12}, we also required $\log_{10}(f_{\rm x}/f_{\rm IR}) \leq -1$.
In addition, we used optical and near-infrared magnitudes to classify foreground stars.
They are expected to be brighter in $R$ than background objects and
with colors $B-R$ and $J-K$ overlapping the intrinsic colours
of main sequence, giant and supergiant stars 
of our Galaxy showed in Fig. \ref{fig. b-r_j-k.ps} with three lines
(obtained from \citealt{Johnson66}) .

Another method to classify Galactic
objects is based on proper motions. 
We used the proper motion measurements provided by the \emph{PPMXL Catalog
of positions and proper motions on the ICRS} \citep{Roeser10} or 
other catalogues of proper motions (e.g. the \emph{Yale/San Juan Southern 
Proper Motion Catalog 4 (SPM4)} \citealt{Girard11} and 
\emph{the Fourth U.S. Naval Observatory CCD Astrograph Catalog (UCAC4)} \citealt{Zacharias13}).

From previous considerations we classified a source to be a foreground star
when all these conditions were met:
\begin{itemize}
\smallskip
\item $\log_{10}(f_{\rm x}/f_{\rm opt}) \leq -1$;
\smallskip
\item $\log_{10}(f_{\rm x}/f_{\rm IR}) \leq -1$;
\smallskip
\item $HR_2 \lesssim 0.3$;
\smallskip
\item $HR_3 \lesssim -0.4$;
\smallskip
\item $J-K_{\rm s} \lesssim 1.0$.
\smallskip
\end{itemize}
or:
\begin{itemize}
\smallskip
\item measurement of the proper motion provided by the PPMXL catalogue.
\smallskip
\end{itemize}

Six sources met these criteria, hence we classified them
as foreground stars (Table \ref{Tab. source-offset list}).
Four of them (No.\,4, 9, 40, 89) show X-ray, optical, and infrared properties typical of 
normal stars. The remaining No.\,13 and 94 have peculiar properties
indicating the possible presence of an accreting companion star.
In particular, both have a radio (ATCA) counterpart.
Source No.\,82 lies on the lines of intrinsic colours of stars in Fig. \ref{fig. b-r_j-k.ps},
but its HR$_3$ is too hard for a foreground star (see Fig. \ref{fig. log10-fx-fopt}).

A detailed discussion of the classification of the most interesting
foreground stars is provided in Sect. \ref{sect. app. fgstars}.

\begin{table*}
\begin{center}
\caption{X-ray sources classified as foreground star candidates 
and the associated optical counterparts. 
pmRA and pmDec are the proper motion of the star in milli-arcsec per year.
Proper motion velocities are taken from the PPMXL catalogue.}
\label{Tab. source-offset list}
\begin{tabular}{lccc@{\, \,}c@{\, \,}c@{\, \,}cc}
\hline
\hline
No.&     RA      &     Dec      &     USNO-B1    & B\,mag.& R\,mag. &      pmRA      &       pmDec      \\
   &   (J2000)   &   (J2000)    &                &        &         & \multicolumn{2}{c}{(mas yr$^{-1}$)}\\
\hline                                                                           
4              & 04 02 59.40 & $-$43 32 23.6  &  0464-0031137  &  15.13 &  14.02  &                &                  \\
9$^{\dagger *}$  & 04 03 07.15 & $-$43 30 29.0  &  0464-0031154  &  12.61 &  11.59  &  $-6.2\pm1.8$  &   $-7.4\pm1.8$   \\
13             & 04 03 12.96 & $-$43 11 57.6  &  0468-0031995  &  21.25 &  19.64  &  $85.4\pm10.3$ &  $-49.8\pm10.3$  \\
40$^{\dagger *}$ & 04 03 40.85 & $-$43 28 57.1  &  0465-0031108  &  12.28 &  11.1   &  $18.3\pm 1.7$ &   $-3.0 \pm 1.7$ \\
89             & 04 04 31.49 & $-$43 26 10.4  &  0465-0031223  &  12.69 &  11.34  &  $12.5\pm 2.1$ &   $9.6 \pm 2.1$ \\
94$^{\dagger}$  & 04 04 43.73 & $-$43 25 22.9  &  0465-0031250  &  14.63 &  14.72  & $-16.3\pm10.4$ &  $10.4 \pm 10.4$  \\
\hline
\end{tabular}
\end{center}
Notes: \\
$^\dagger$: this star also has an independent proper motion measurement in the Yale/San Juan Southern Proper Motion Catalog 4 (SPM4) \citep{Girard11};
$^*$: this star also has an independent proper motion measurement in the Fourth U.S. Naval Observatory CCD Astrograph Catalog (UCAC4) \citep{Zacharias13}.
\end{table*}

\subsection{Background objects}
\label{sect. bg objects}

Another class of sources included in the X-ray observations of nearby galaxies
are the background objects: normal galaxies, AGNs, and galaxy clusters.

The identification of a background object in this work 
is based on SIMBAD and NED correlations. 
This allowed us to identify source No.\,61 with a galaxy
(MRSS~250-126782, \citealt{Ungruhe03},
and 2MASX~J04035631-4329046, 2 Micron All Sky Survey Extended objects, 
\citealt{Skrutskie06}; see Table \ref{Tab. list-galaxies}).

The existence of a radio counterpart
and a hardness ratio of $HR_2 \geq -0.4$ \citep{Pietsch04} 
are indicative of an AGN. 
We found radio counterparts for the hard sources No.\,23, 67, 71, 83, 98 
(see Table \ref{Tab. list-galaxies} and Sect. \ref{sect. app. bg-obj})
and classified them as AGNs for the first time.

\begin{table}
\begin{center}
\caption{X-ray sources identified and classified as galaxies or AGNs.}
\label{Tab. list-galaxies}
\resizebox{\columnwidth}{!}{
\begin{tabular}{lcccccc}
\hline
\hline
    &\multicolumn{2}{c}{\emph{XMM-Newton}} & \multicolumn{2}{c}{Counterpart} \\
No. &      RA     &      Dec     & \multicolumn{2}{c}{Name or RA/Dec (J2000)} \\
    &    (J2000)  &    (J2000)   &                    &                       \\
\hline
\multicolumn{5}{c}{Identifications:}\\
\hline                                               
61    & 04 03 56.60 & $-$43 29 05.0 & \multicolumn{2}{c}{2MASS J04035632-4329041} \\
\hline
\multicolumn{5}{c}{Classification:}\\
\hline                                               
23$^\dagger$& 04 03 25.37 & $-$43 17 22.0 & 04 03 25.54 & $-$43 17 23.2 \\
67$^\dagger$& 04 04 08.90 & $-$43 18 43.8 & 04 04 08.82 & $-$43 18 42.3 \\
71$^\dagger$& 04 04 12.64 & $-$43 22 30.7 & \multicolumn{2}{c}{SUMSS J040412-432235} \\ 
83$^\dagger$& 04 04 22.57 & $-$43 24 52.3 & 04 04 22.25 & $-$43 24 52.1 \\
98$^\dagger$& 04 04 47.14 & $-$43 22 03.5 & 04 04 46.77 & $-$43 22 02.2 \\
\hline
\end{tabular}
}
\end{center}
Notes: \\
$^\dagger$ radio sources detected by ATCA
\end{table}

\begin{table*}
\begin{center}
\caption{Best-fitting parameters of the X-ray spectra of sources No.\,23, 54, and 63 (errors at 90\% confidence level).
The fluxes (erg cm$^{-2}$ s$^{-1}$) have been calculated in the energy range 
$0.2-10$ keV. Galactic absorption: $N_{\rm H}=10^{20}$~cm$^{-2}$.}
\label{Tab. no.23 54 63 spectral parameters}
\begin{tabular}{lcccccccc}
\hline
\hline
\noalign{\smallskip}
No.&         $N_{\rm H}$      &         $\Gamma$        & $kT$ (\texttt{diskbb}) & \multicolumn{2}{c}{$\chi^2_\nu$ (d.o.f.)$^\dagger$}&              $F_{\rm x}$            &           unabs. $F_{\rm x}$          \\
   &   ($10^{21}$\,cm$^{-2}$) &                         &          (keV)         & c-stat (d.o.f.) & goodness &  \multicolumn{2}{c}{($0.2-10$ keV) erg cm$^{-2}$ s$^{-1}$}                   \\
\noalign{\smallskip}
\hline 
\noalign{\smallskip}
23 & $2.8{+0.9 \atop -0.7}$ &$3.6{+0.5 \atop -0.4}$   &                        &\multicolumn{2}{c}{1.25 (31)}& $6.9{+6.1 \atop -3.2}\times 10^{-14}$& $7.0{+5.7 \atop -3.4}\times 10^{-13}$\\
\noalign{\smallskip}
30 & $\lesssim 0.1$          &                        &$0.66{+0.17 \atop -0.15}$& 367.18 (442) & 51.19\% & $1.9{+0.4 \atop -0.4}\times 10^{-14}$& $1.7{+0.4 \atop -0.4}\times 10^{-14}$\\
\noalign{\smallskip}
33 &$0.04{+0.44 \atop -0.04}$&                       &$1.26{+0.37 \atop -0.30}$& 248.92 (365) & 2.97\%  & $2.7{+0.7 \atop -0.6}\times 10^{-14}$& $2.8{+13.0 \atop -1.8}\times 10^{-14}$\\
\noalign{\smallskip}
54 & $\lesssim 0.2$          & $1.4{+0.4 \atop -0.4}$ &$0.28{+0.03 \atop -0.04}$  &\multicolumn{2}{c}{1.10 (65)}& $8.8{+0.9 \atop -0.9}\times 10^{-14}$& $9.3{+1.0 \atop -0.8}\times 10^{-14}$\\
\noalign{\smallskip}
58 &$0.02{+0.26 \atop -0.02}$& $2.0{+0.3 \atop -0.2}$&                         & 453.52 (546) & 13.11\% & $3.4{+0.6 \atop -0.5}\times 10^{-14}$& $1.6{+2.1 \atop -0.8}\times 10^{-13}$\\
\noalign{\smallskip}
63 & $\lesssim 0.1$          &$2.2{+0.2 \atop -0.1}$  &                        &\multicolumn{2}{c}{0.90 (29)}& $9.4{+0.7 \atop -1.1}\times 10^{-14}$& $1.0{+0.2 \atop -0.5}\times 10^{-13}$\\
\noalign{\smallskip}
92 & $\lesssim 0.1$          &                         &$0.58{+0.13 \atop -0.10}$& 321.24 (413) & 23.16\% & $4.5{+0.8 \atop -0.7}\times 10^{-14}$& $4.4{+0.7 \atop -0.7}\times 10^{-14}$\\
\noalign{\smallskip}
\hline
\end{tabular}
\end{center}
Notes: \\
$^\dagger$ Reduced $\chi^2$ and d.o.f. or Cash statistic, d.o.f. and percentage of realisations with statistic $<$ c-stat
\end{table*}

\begin{figure*}
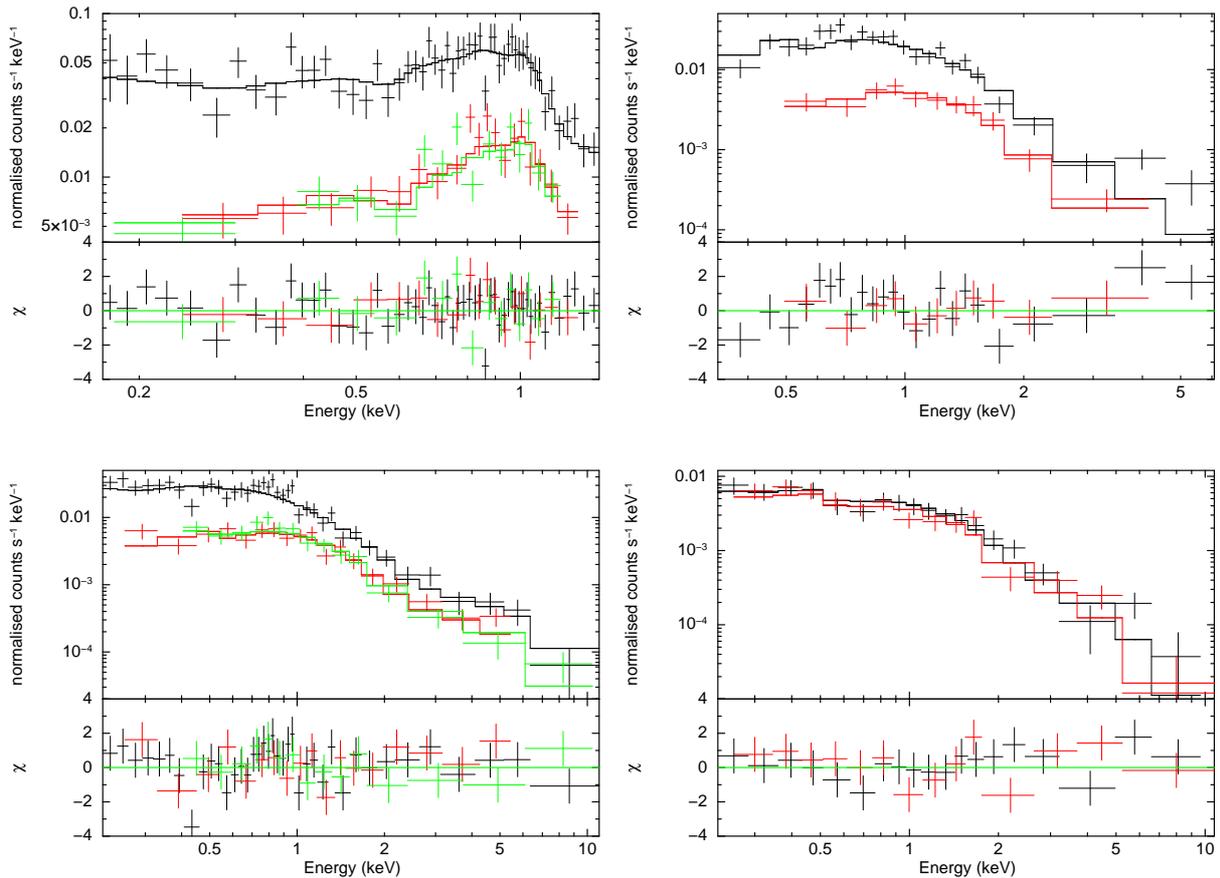

\begin{center}
\includegraphics[bb=25 -21 583 720, clip, angle=-90,width=8cm]{s9_mekal_2T_abs.ps}
\includegraphics[bb=25 -21 583 720, clip, angle=-90,width=8cm]{s23_po.ps}
\includegraphics[bb=25 -21 583 720, clip, angle=-90,width=8cm]{s54_diskbb_po.ps}
\includegraphics[bb=25 -21 583 720, clip, angle=-90,width=8cm]{s63_po_new.ps}
\end{center}
\caption{EPIC counts spectra, together with residuals in units of standard deviations
for sources No.\,9 (top-left panel), No.\,23 (top-right panel; PN: black; MOS2: red), 
No.\,54 (bottom-left panel; PN: black; MOS1: red; MOS2: green), and No.\,63 (bottom-right panel; MOS1: black; MOS2: red).}
\label{fig. no.9 23 54 63}
\end{figure*}

\subsection{Sources in NGC 1512/1510}
\label{sect. sources in NGC 1512/1510}

We detected 21 sources inside the $D_{25}$ ellipses
(NGC 1512: $8.9\times 5.6$ arcmin$^2$; NGC 1510: $1.3\times 0.7$ arcmin$^2$; \citealt{deVaucouleurs91})
of NGC 1512 and NGC1510 or overlapping the outer galaxy arms detected by GALEX.
These sources, with their positions, hardness ratios, fluxes and luminosities (assuming $d=9.5$~Mpc)
are listed in Table \ref{Tab. sources in ngc1512-1510}.
The detection-limiting flux in the $0.2-12$~keV energy band is 
$\sim 2.6 \times 10^{-15}$~erg~cm$^{-2}$~s$^{-1}$ for sources inside the $D_{25}$ ellipses.
In Table \ref{Tab. sources in ngc1512-1510}
we did not include sources identified and classified as background objects
or foreground stars. 
We included source No.\,63, classified as quasi-stellar object (QSO) in the catalogue of \citet{Atlee07},
whose intermediate ultra-luminous X-ray source (ULX) nature cannot be ruled out (see Sect. \ref{sect. remaining sources}).
Sources No.\,54 and 29 correspond to the nuclear regions of NGC\,1512 and NGC\,1510.
The properties of source No.\,54 are discussed in Sect. \ref{sect. nuclear regions}.
The remaining 20 sources are discussed in Sects. \ref{sect. source No. 63} and \ref{sect. remaining sources}.

\section{Discussion}
\label{sect. discussion}

\subsection{Nuclear source of NGC~1512 (XMMU~J040354.2-432056)} 
\label{sect. nuclear regions}

Source No.\,54 (XMMU~J040354.2-432056) is the nuclear source of NGC~1512 and
is the brightest source inside the $D_{25}$ ellipse.
Because of the relatively poor spatial resolution of \xmm,
we cannot study this region of NGC~1512 in detail,
as the X-ray emission is expected to be produced by diffuse emission,
unresolved point sources (dominated by XRBs), and a possible active nucleus 
with a low accretion rate in the galactic centre.

The $0.2-10$~keV \xmm\ light curve and the \sw/XRT measurements of the flux (see Table \ref{Tab. variability})
show neither short- nor long-term variability.
The source was bright enough to allow spectral analysis.
We extracted PN, MOS1, and MOS2 spectra and fitted them simultaneously 
using different spectral models.
The best-fit model ($\chi^2_\nu=1.08$, 66 d.o.f.) is an absorbed power-law plus disc-blackbody,
with a photon-index of $\Gamma = 1.4 \pm 0.4$, a temperature of
$kT_{\rm in}=0.28{+0.03 \atop -0.04}$~keV, and a luminosity of
$L_{\rm x} = 1.00{+0.11 \atop -0.09} \times 10^{39}$ erg s$^{-1}$ in the energy band $0.2-10$~keV
(Table \ref{Tab. no.23 54 63 spectral parameters}; Fig. \ref{fig. no.9 23 54 63}).
The spectral parameters and the X-ray luminosity are typical of accreting 
intermediate-mass black holes (IMBHs; see e.g. \citealt{Stobbart06}).
We also obtained an acceptable fit ($\chi^2=0.90$, 65 d.o.f.)
with an absorbed power-law plus thermal component, \texttt{phabs(powerlaw+apec)},
with $\Gamma = 2.10 \pm 0.11$ and $kT = 0.58{+0.13 \atop -0.22}$~keV.
Although this spectral model overfits the data with a $\chi^2=0.90$, 
it describes the expected X-ray emission from the unresolved nuclear region of a galaxy better,
where a fraction of the X-ray emission is produced by point sources
and the remaining emission comes from interstellar gas.
The power-law photon index obtained with this fit is compatible with the spectral emission from
an active nucleus with a very low accretion rate\footnote{Photon indices 
in the range of $\Gamma \sim 1.5-2.5$ have been measured in different AGNs 
(\citealt{Ishibashi10} and references therein).},
or with the spectral emission from XRBs.

\subsection{Source No.\,63 (XMMU~J040400.9-432319)}
\label{sect. source No. 63}

The second-brightest source in the $D_{25}$ of NGC~1512 is source No.\,63.
It coincides with the infrared source 2MASS J04040093-4323175
and the optical source USNO-B1 0466-0031610.
We extracted the MOS1 and MOS2 spectra (the position of the source
was close to a gap in PN) and found that an absorbed power-law
provides an acceptable fit (see Table \ref{Tab. no.23 54 63 spectral parameters} 
and Fig. \ref{fig. no.9 23 54 63}). No.\,63 does not show any significant variability in the
\xmm\ and \sw\ observations (Table \ref{Tab. variability}).
\citet{Atlee07} classified the optical counterpart of No.\,63 as a QSO candidate 
according to the selection criteria based on optical (USNO~A2.0, \citealt{Monet98})
and UV (GALEX Release Two) properties. The catalogue of \citet{Atlee07}
provides a probability of 36\% that this source is a QSO candidate.
Therefore we investigated other possible classifications.
A comparison between the optical/infrared colours and the optical luminosity
from USNO~B1 and 2MASS catalogues with the stellar spectral flux library of \citet{Pickles98}
shows that No.\,63 cannot be a foreground star.
If No.\,63 belongs to NGC~1512, its $0.5-10$~keV luminosity would be
$8-13\times 10^{38}$~erg~s$^{-1}$, exceeding the Eddington limit of an
accreting neutron star and compatible with an accreting black hole
with an X-ray luminosity lying just below the conventional luminosity threshold
of $\sim 10^{39}$~erg~s$^{-1}$ of ULXs.

\subsection{Other sources in NGC\,1512/1510} 
\label{sect. remaining sources}

The $0.2-12$~keV fluxes of the remaining 19 sources range from $\sim 3.4 \times 10^{-15}$~erg~cm$^{-2}$~s$^{-1}$
to $\sim 9.6 \times 10^{-14}$~erg~cm$^{-2}$~s$^{-1}$ ($3.7 \times 10^{37} \lesssim L_x \lesssim 10^{39}$~erg~s$^{-1}$,
assuming $d=9.5$~Mpc). These sources are too faint to perform a spectral analysis.
Therefore we used the hardness ratios to outline their X-ray properties.
Figure \ref{fig. hr diagrams} shows the hardness ratios of sources in NGC\,1512/1510
with uncertainties less than 0.3 and a grid of hardness ratios 
calculated for different spectral models (Sect. \ref{sect. spectral analysis and hardness-ratios}).
The entire sample of sources plotted in Fig. \ref{fig. hr diagrams}
shows hardness ratios compatible with power-law or disc-blackbody models,
indicating that they might either be accreting compact objects in
NGC\,1512/1510 or background objects.

\begin{table}
\begin{center}
\caption{\xmm\ sources within the $D_{25}$ ellipses of NGC 1512 and NGC 1510.}
\label{Tab. sources in ngc1512-1510}
\resizebox{\columnwidth}{!}{
\begin{tabular}{lcclcr@{}l}
\hline
\hline
 \noalign{\smallskip}
No. &       RA    &      Dec    &\multicolumn{1}{c}{Flux ($0.2-12$ keV)}&\multicolumn{3}{c}{Luminosity ($0.2-12$ keV)}\\
    &   (J2000)   &   (J2000)   &\multicolumn{1}{c}{erg cm$^{-2}$ s$^{-1}$}&       \multicolumn{3}{c}{erg s$^{-1}$} \\
 \noalign{\smallskip}
\hline
 \noalign{\smallskip}
24$^1$& 04 03 25.76 & $-$43 23 10.0 & $(1.7 \pm 0.5)\times 10^{-14}$ &\ \ &$(1.8 \pm 0.5)$ & $\times 10^{38}$\\ 
 \noalign{\smallskip}
28    & 04 03 32.12 & $-$43 21 39.2 & $(1.9 \pm 0.2)\times 10^{-14}$ &\ \ &$(2.1 \pm 0.2)$ & $\times 10^{38}$\\ 
 \noalign{\smallskip}
29$^2$& 04 03 32.71 & $-$43 24 01.7 & $(2.3 \pm 0.6)\times 10^{-14}$ &\ \ &$(2.5 \pm 0.6)$ & $\times 10^{38}$\\ 
 \noalign{\smallskip}
42    & 04 03 42.10 & $-$43 18 38.5 & $(8.4 \pm 3.4)\times 10^{-15}$ &\ \ &$(9.1 \pm 3.7)$ & $\times 10^{37}$\\ 
 \noalign{\smallskip}
43    & 04 03 42.40 & $-$43 19 58.9 & $(1.6 \pm 0.4)\times 10^{-14}$ &\ \ &$(1.7 \pm 0.4)$ & $\times 10^{38}$\\ 
 \noalign{\smallskip}
44    & 04 03 42.72 & $-$43 23 17.1 & $(1.1 \pm 0.3)\times 10^{-14}$ &\ \ &$(1.2 \pm 0.3)$ & $\times 10^{38}$\\ 
 \noalign{\smallskip}
45    & 04 03 42.90 & $-$43 20 50.2 & $(5.2 \pm 2.7)\times 10^{-15}$ &\ \ &$(5.6 \pm 2.9)$ & $\times 10^{37}$\\ 
 \noalign{\smallskip}
51    & 04 03 50.86 & $-$43 22 04.9 & $(5.1 \pm 2.5)\times 10^{-15}$ &\ \ &$(5.5 \pm 2.7)$ & $\times 10^{37}$\\ 
 \noalign{\smallskip}
53    & 04 03 52.76 & $-$43 20 59.0 & $(4.3 \pm 3.2)\times 10^{-15}$ &\ \ &$(4.6 \pm 3.4)$ & $\times 10^{37}$\\ 
 \noalign{\smallskip}
54$^3$& 04 03 54.26 & $-$43 20 56.7 & $(9.8 \pm 0.6)\times 10^{-14}$ &\ \ &$(10.6 \pm 0.6)$ & $\times 10^{38}$\\ 
 \noalign{\smallskip}
55    & 04 03 54.30 & $-$43 18 53.0 & $(7.6 \pm 2.6)\times 10^{-15}$ &\ \ &$(8.2 \pm 2.8)$ & $\times 10^{37}$\\ 
 \noalign{\smallskip}
57    & 04 03 54.79 & $-$43 22 24.5 & $(6.7 \pm 5.0)\times 10^{-15}$ &\ \ &$(7.2 \pm 5.4)$ & $\times 10^{37}$\\ 
 \noalign{\smallskip}
59$^1$& 04 03 56.14 & $-$43 17 20.7 & $(1.6 \pm 0.3)\times 10^{-14}$ &\ \ &$(1.7 \pm 0.3)$ & $\times 10^{38}$\\ 
 \noalign{\smallskip}
62    & 04 03 58.34 & $-$43 22 46.0 & $(1.5 \pm 0.5)\times 10^{-14}$ &\ \ &$(1.6 \pm 0.5)$ & $\times 10^{38}$\\  
 \noalign{\smallskip}
63$^4$& 04 04 00.99 & $-$43 23 19.0 & $(9.6 \pm 0.7)\times 10^{-14}$ &\ \ &$(10.4 \pm 0.8)$ & $\times 10^{38}$\\
 \noalign{\smallskip}
67    & 04 04 08.90 & $-$43 18 43.8 & $(3.4 \pm 1.7)\times 10^{-15}$ &\ \ &$(3.7 \pm 1.8)$ & $\times 10^{37}$\\
 \noalign{\smallskip}
69    & 04 04 11.17 & $-$43 19 57.7 & $(1.6 \pm 0.4)\times 10^{-14}$ &\ \ &$(1.7 \pm 0.4)$ & $\times 10^{38}$\\
 \noalign{\smallskip}
73$^1$& 04 04 14.97 & $-$43 22 57.6 & $(1.6 \pm 0.5)\times 10^{-14}$ &\ \ &$(1.7 \pm 0.5)$ & $\times 10^{38}$\\
 \noalign{\smallskip}
74$^1$& 04 04 15.17 & $-$43 22 34.3 & $(4.1 \pm 3.4)\times 10^{-15}$ &\ \ &$(4.4 \pm 3.6)$ & $\times 10^{37}$\\
 \noalign{\smallskip}
75$^1$& 04 04 15.21 & $-$43 24 09.8 & $(4.0 \pm 3.9)\times 10^{-15}$ &\ \ &$(4.3 \pm 4.2)$ & $\times 10^{37}$\\
 \noalign{\smallskip}
90$^1$& 04 04 35.29 & $-$43 20 06.7 & $(8.5 \pm 7.0)\times 10^{-15}$ &\ \ &$(9.2 \pm 7.5)$ & $\times 10^{37}$\\
 \noalign{\smallskip}
\hline
\end{tabular}
}
\end{center}
Notes: $^1$ outside of $D_{25}$ but overlapping outer galaxy arms of GALEX observation;
$^2$ nuclear region of NGC 1510;
$^3$ nuclear region of NGC 1512;
$^4$ QSO or intermediate ULX;
\end{table}

\xmm\ detected 13 non-nuclear sources inside the $D_{25}$ ellipse of NGC~1512 in the energy band $0.2-12$~keV.
Based on the spectral emission properties emerging from the hardness ratios, they can be either XRBs or AGNs.
Assuming that AGNs are uniformly distributed in the sky, their number within the $D_{25}$ ellipse of NGC~1512
can be derived statistically using the $0.5-2$~keV and $2-10$~keV
$\log N - \log S$ AGN distributions of \citet{Cappelluti09}.
The $\log N - \log S$ in the energy range $2-10$~keV is almost unaffected by the incompleteness effect
produced by the absorption. In contrast, this effect is present in the energy range $0.5-2$~keV.
We accounted for NGC~1512 absorption in the $0.5-2$~keV $\log N - \log S$ 
by assuming that the total column density through the galactic disc (along the directions perpendicular
to the galactic plane) is $5\times 10^{20}$~cm$^{-2}$,
in agreement with the hydrogen distribution 
in the direction perpendicular to the Galactic plane of the Milky Way \citep{Kalberla09}.
Then, assuming that the spectral emission of AGNs in $0.5-2$~keV
is described by a power-law with photon index $\Gamma=2$ (see \citealt{Cappelluti09} and references therein),
we found that AGNs beyond NGC~1512 detected with a flux $\geq 1.1 \times 10^{-15}$~erg~cm$^{-2}$~s$^{-1}$ 
would have fluxes $\geq 1.2 \times 10^{-15}$~erg~cm$^{-2}$~s$^{-1}$ 
if they were not absorbed by NGC~1512.
We found that $7 \pm 0.5$ AGNs are expected in the $D_{25}$ ellipse of NGC~1512 in the energy range $0.5-2$~keV
with observed fluxes $\geq 1.1 \times 10^{-15}$~erg~cm$^{-2}$~s$^{-1}$
and $6.9 \pm 0.5$ AGNs in the energy range $2-10$~keV
with observed fluxes higher than the $2-10$~keV limiting flux $4.3 \times 10^{-15}$~erg~cm$^{-2}$~s$^{-1}$.
The numbers derived above can be compared with the numbers of sources
detected by \xmm\ with a higher detection likelihood than 6
in the energy bands $0.5-2$~keV and $2-10$~keV, 
within the $D_{25}$ ellipse of NGC~1512.
To accomplish this, we corrected the number of X-ray sources
for the incompleteness effect.
It consists of an underestimation of the number of sources detected
in the surveyed area caused by the non-uniformity of the
sensitivity of the EPIC instruments across the field of view.
We corrected the number of X-ray sources in the $D_{25}$ region as follows:
we first created the combined sensitivity maps of the three EPIC instruments
in the energy bands of $0.5-2$~keV and $2-10$~keV with the SAS task \texttt{esensmap}.
We excluded a circular region centred on the nuclear region of NGC~1512
with radius 12~arcsec from this analysis, where the relatively poor spatial resolution of EPIC
in a crowded region causes source confusion effects.
$A_{\rm tot}$ is the area of the sky surveyed by EPIC.
We used the resulting sensitivity maps to calculate 
the fraction of surveyed area $\Omega(F_i)/A_{\rm tot}$, 
over which sources with a given flux $F_i$
can be detected. Then, we used this factor to correct the total number
of detected sources for each of the two energy bands considered for incompleteness
with the formula $N=\sum_{i=1}^{N_{\rm d}}A_{\rm tot}/\Omega(F_i)$,
where $N_{\rm d}$ is the number of detected sources
and $N$ is the number of sources corrected for incompleteness.
We found seven sources with a flux higher than $1.1 \times 10^{-15}$~erg~cm$^{-2}$~s$^{-1}$ 
in the energy range of $0.5-2$~keV and nine sources with a flux higher than 
$4.3 \times 10^{-15}$~erg~cm$^{-2}$~s$^{-1}$ in the energy range $2-10$~keV.
From a simple comparison we find that the number of expected AGNs calculated above
agrees; these numbers are slightly lower than the number of 
X-ray sources observed by \xmm\ and corrected for incompleteness, 
which means that they indicate a small population of XRBs in NGC~1512.

Since the number of HMXBs and LMXBs in nearby galaxies
depends on the SFR and the stellar mass content, respectively
(see e.g. \citealt{Grimm03}; \citealt{Gilfanov04}),
we obtained a rough estimate of the expected number of HMXBs and LMXBs in the $D_{25}$ ellipse of NGC~1512 
above the limiting flux of $4.3 \times 10^{-15}$~erg~cm$^{-2}$~s$^{-1}$ in the energy range of $2-10$~keV.
Using the $N_{\rm HMXB}-SFR$ relation of \citet{Mineo12} (eq. 18)
and $SFR \sim 0.2$~M$_\odot$~yr$^{-1}$ \citep{Koribalski09}
we found $N_{\rm HMXB} \approx 1$.
The number of LMXBs depends on the stellar mass $M_*$ of a galaxy.
We obtained $M_* \approx 10^{10}$~M$_\odot$ using the relation 
between the mass-to-light ratio $M_*/L_K$ and the $B-V$ optical colour
of \citet{Bell01} ($\log [M_*/L_{\rm K}] = -0.692 +0.652[B-V]$), 
the extinction-corrected total extrapolated magnitude $K_s$ from the 2MASS redshift catalogue
\citep{Huchra12}, and the extinction-corrected $B-V$ optical colour from \citet{deVaucouleurs91} RC3 catalogue.
Using the X-ray luminosity function calculated by \citet{Gilfanov04} (eq. 8),
we obtained $N_{\rm LMXB} \approx 4$ in NGC~1512. 
The sum of the number of HMXBs and LMXBs obtained from SFR and $M_*$ measurements and
the number of expected AGNs agree with the 
number of X-ray sources observed by \xmm\ within the $D_{25}$ ellipse of NGC~1512.

\subsection{Diffuse emission}
\label{sect. diffuse emission}

We have detected a soft X-ray emitting component that is located in the ring of \Gal. 
Assuming a thermal plasma model for the emission yields a temperature of $kT=0.66~(0.28-0.89)$~keV. 
Typically, thermal emission from spiral galaxies exhibits a two-temperature spectrum with cool 
and hot components of $kT_{\rm cool}\sim0.2$~keV and $kT_{\rm hot}\sim0.7$~keV. For example, this two temperature model 
has been observed in M101 \citep{Kuntz2003}, M81 \citep{Page2003}, and in the barred-spirals 
NGC~6946 \citep{Schlegel2003} and NGC~1672 \citep{Jenkins2011}. The temperature of our best-fit spectral model to 
the inner \Gal\ spectrum agrees with the hotter component expected for spiral galaxies. We suspect that, 
because of the poor count statistics and significant contamination of our spectra, we simply lack the sensitivity 
to detect the expected soft component. Accordingly, we considered only the detected hot component but caution that 
it may not represent the underlying X-ray emission from the centre of \Gal. The detected emission may not 
be truly diffuse in origin. Unresolved discrete sources can contribute to the apparently diffuse source and must therefore be quantified. 

\subsubsection{Unresolved HMXBs}
\par Typically, one can separate the unresolved discrete sources into contributions from objects associated 
with the young (HMXBs) and older (LMXBs; cataclysmic variables, CVs; and coronally active binaries, ABs) 
stellar populations in the region. \citet{Koribalski09} determined the ages of the UV-bright stellar populations 
in the nuclear region and ring of \Gal\  and found ages $>130$~Myr. Therefore, it is unlikely that any HMXBs 
are located in the central regions because they are associated with younger stellar populations with ages $<100$~Myr.

\subsubsection{Unresolved weak sources}
Objects associated with the older stellar population are much more likely to contribute. \citet{Rev2007, Rev2008} 
demonstrated that the X-ray luminosity per unit stellar mass due to the weak X-ray sources (CVs and ABs)
associated with the old stellar population in the solar vicinity \citep{Saz2006} is compatible 
with that observed in other galaxies. Hence, we can determine the expected X-ray luminosity from the unresolved 
CVs and ABs using the mass or, equivalently, luminosity of the stellar population, which can be estimated 
using NIR data (see also Sect. \ref{sect. remaining sources}). We determined the combined $K_{s}$-band luminosity 
of the central region of \Gal\ using the 2MASS Large Galaxy Atlas \citep[][see Fig. \ref{soft-inner-ngc1512}~right]{Jarrett2003}. 
Conveniently, the isophotal $K_{s}$-band magnitude in the atlas is representative of the central region with $K_{s}=7.7$. 
Converting this into a luminosity gives $L_{K_{s}} = 2.1\times10^{9}$~L$_{\sun}$. We then used the relation 
$L_{\rm{X,0.5-2~keV}}/L_{K_{s}} = (5.9 \pm 2.5) \times 10^{27}$~erg~s$^{-1}$~$L^{-1}_{K_{s},\rm{L}_{\sun}}$ \citep{Rev2008} 
and adjusted for the X-ray energy range using their emission model and the 
WebPIMMS\footnote{Available at \burl{https://heasarc.gsfc.nasa.gov/Tools/w3pimms.html}} 
tool to determine an expected contribution to the X-ray luminosity of $L_{\rm{X,~0.3-10~keV}} = (1.9 \pm 0.8) \times 10^{37}$~erg~s$^{-1}$.

\subsubsection{Unresolved LMXBs}
\par Unresolved LMXBs associated with the old stellar population can also contribute to the apparently diffuse emission. 
Since the number of LMXBs in a stellar population is related to the total mass \citep{Gilfanov04}, 
we estimated the X-ray luminosity of LMXBs below the detection threshold using the total stellar mass in the central region. 
Using the method outlined in Sect. \ref{sect. remaining sources} with the combined $K_{s}$-band magnitude of the stellar population, 
we estimated the total stellar mass in the central region to be $\sim1.3 \times 10^{9}$~M$_{\sun}$. 
We then used Equation 11 of \citet{Gilfanov04} to estimate the number of LMXBs with 
$L_{\rm{X}}>10^{37}$~erg~s$^{1}$ and their combined luminosity to be $\sim2$ and $1.0 \times 10^{38}$~erg~s$^{-1}$, respectively. 
As stated by \citet{Gilfanov04}, this X-ray luminosity is related to the total $L_{\rm{X}}$ due to LMXBs
through a correction factor of $\approx1.1$. Thus, the total LMXB population in the central region contributes 
$L_{\rm{X}} = 1.1 \times 10^{38}$~erg~s$^{1}$. In addition to the nuclear source, there are two X-ray sources located 
in the interior region (see Fig. \ref{fig. xmm-mosaic}). The X-ray colours of these sources suggest that they are either 
XRBs or AGN, namely No.\,51 and No.\,53 (see Sect. \ref{sect. remaining sources}). The stellar mass in the central region suggests 
that probably about two sources with $L_{\rm{X}}>10^{37}$~erg~s$^{1}$ are present. Since our source detection 
sensitivity is $\gtrsim  10^{37}$~erg~s$^{1}$, No.\,51 and No.\,53 might be expected to be LMXBs. 
If that is the case, their combined $L_{\rm{X}} = (9.6 \pm 4.3) \times 10^{37}$~erg~s$^{1}$,
which is similar to the total emission expected from the LMXB population. 
This means that if these are indeed LMXBs, 
the $L_{\rm{X}}$ due to unresolved LMXBs is  $< 5.7 \times 10^{37}$~erg~s$^{1}$, 
with the upper limit being about half of the observed $L_{\rm{X}}$ in the central region. \\

\subsubsection{Emission from the interstellar gas}
With these unresolved source contributions in mind, we re-ran our spectral fits, this time including 
spectral components representing the unresolved discrete sources, 
to determine the plasma temperature and luminosity of the diffuse emission.
We fixed the normalisations of these additional 
components so that they represented the upper limits of their luminosity contribution. For the CVs and ABs we used
the model of \citet{Rev2008}, a combination of a thermal plasma (\texttt{mekal} in \texttt{XSPEC}) with solar abundance 
and $kT = 0.5$~keV, and a power-law with photon index $\Gamma = 1.9$. For the LMXBs, we used a power-law model with 
photon index $\Gamma = 1.5$, typical of LMXBs \citep{Soria2003}. The resulting plasma temperature for the truly 
diffuse emission is $kT=0.68~(0.31-0.87)$~keV and $L_{\rm{X, 0.3-10~\rm{keV}}} = 1.3 \times 10^{38}$~erg~s$^{-1}$. 
\citet{Mineo2012} determined from a sample of nearby star-forming galaxies that the $0.5-2$~keV luminosity 
of diffuse X-ray emission from the ISM is linearly correlated with the SFR of that galaxy as 
$L_{X, 0.5-2~\rm{keV}} (\rm{erg~s}^{-1}) \approx (8.3 \pm 0.1) \times 10^{38} \rm{SFR~(M}_{\sun}\rm{yr}^{-1})$. 
Because the SFR of \Gal\  is 0.22~M$_{\sun}$~yr$^{-1}$ \citep{Koribalski09}, 
we expect $L_{\rm{X, 0.5-2~\rm{keV}}}$ to be $\sim2 \times 10^{38}$~erg~s$^{-1}$. 
The observed $L_{\rm{X, 0.5-2~\rm{keV}}}$ was found to be $\sim1\times 10^{38}$~erg~s$^{-1}$, 
which is largely consistent with the relation of \citet{Mineo2012}. 
An explanation of the slight discrepancy may be the non-detection of the cooler ISM component, 
typically observed in star-forming galaxies, 
because of the sensitivity of our observation (see Sect. \ref{sect. diffuse emission}).

In the inner regions of spiral galaxies, there are three possible sources of a hot X-ray emitting gas:
stellar winds and supernova remnants associated with the older stellar population, superwinds driven by 
active star formation, and an AGN \citep{Tyler2004}. Since the youngest stellar population 
in the inner \Gal\ region is of intermediate age ($\sim130$~Myr), it is unlikely that a superwind is being driven 
as the massive stellar population has been exhausted. Similarly, the absence of an AGN in \Gal\ naturally rules out 
this possible source. Thus, the hot gas is most likely being produced by supernovae from the older stellar population, 
that is, the supernovae currently feeding the hot gas are Type Ia remnants 
because there is no young stellar population ($<100$~Myr) 
capable of producing core-collapse (CC) explosions. However, in the past the contribution from CC remnants 
most likely dominated the energy input because of the intense star formation in the ring and nuclear region 
of \Gal\ caused by the interaction with NGC~1510, very likely around $\sim400$~Myr ago \citep{Koribalski09}. 
The combined effect of the massive stellar winds and supernovae may even have driven a superwind, 
that blew hot gas out along the poles of the galaxy into the halo, entrained and shocked cooler gas 
in the disc and halo, and produced X-rays \citep[e.g.,][]{Strickland2009}. If the radiative cooling time 
were long enough, some contribution from this earlier epoch would be expected. Spectral signatures of a 
Type Ia origin would be evident as enhanced Fe lines in the spectrum. 
However, our spectra lack the count statistics to identify such a signature.

We can estimate some constraints on the physical parameters of the hot diffuse gas based on the results 
from our spectral fits. The normalisation of the thermal plasma model ($K$) representing 
the diffuse component can be related to the emission measure through
\begin{equation}
\label{em}
K = \frac{10^{-14}}{4\pi D^{2}} n_{e}n_{\rm{H}}V .
\end{equation}
For a thermal plasma with 0.65 solar abundance, as is the case for \Gal, $n_{e}/n_{\rm{H}} \sim 1.2$. $V = Ah$, 
where A is the area of the extraction region in cm and $h$ is the height of the emitting region. 
We scaled the $h$ term to units of kpc to allow a direct comparison of our results with other works. 
We also introduced a volume filling factor $f$ to account for the distribution of hot gas in the volume $V$.  
Equation \ref{em} can now be written
\begin{equation}
\label{em2}
n_{e}^{2} = 2.7 \times 10^{-8} \frac{K 4\pi D^{2}}{A f h_{\rm{kpc}}} \rm{cm}^{-6}.
\end{equation}
From our elliptical extraction region with semi-major and semi-minor axes of $\sim1.8\arcmin$ 
and $\sim1.4\arcmin$ (corresponding to $\sim4.9$~kpc and $\sim3.8$~kpc at the distance of \Gal), 
we determined $A$ to be $\sim5.5 \times 10^{44}$~cm$^{2}$. Solving Equation \ref{em2} 
with these values gives $n_{e} = (0.001-0.002) f^{-1/2}h_{\rm{kpc}}^{-1/2}$~cm$^{-3}$. 
Using this determined electron density and the best-fit plasma temperature, 
we also calculated the radiative cooling time ($t_{\rm cool}$) of the gas as
\begin{equation}
\label{tcool}
t_{\rm cool} \sim \frac{3kT}{\Lambda n_{e}},
\end{equation}
with $\Lambda \sim 2 \times 10^{-23}$~erg~cm$^{-3}$~s$^{-1}$ \citep{Suth1993}. 
This yields $t_{\rm cool} \sim 10^{10} f^{1/2}h_{\rm{kpc}}^{1/2}$~yr. 
The determined values of $n_{e}$ and $t_{\rm cool}$ are very low and high, respectively, 
compared with those determined for the central regions of other galaxies. 
\citet{Tyler2004} determined $n_{e}$ to be in the range $0.016-0.11 f^{-1/2}h_{\rm{kpc}}^{-1/2}$~cm$^{-3}$ 
for their sample of spiral galaxies without an AGN. Their higher gas densities naturally lead to lower 
radiative cooling times $\sim 10^{7} f^{1/2}h_{\rm{kpc}}^{1/2}$~yr. 
A possible explanation is that the bar in \Gal\ may play some significant role since bars are known 
to drive gas from the disk to the nuclear regions of a galaxy, fuelling star formation, and feeding an AGN, 
if present \citep{Shlos1990}. Therefore, this dynamic system might have disrupted the hot gas content 
in the inner region of \Gal, resulting in the low electron density we observe. For the moment, 
this is just a suggestion because our data do not allow for a more detailed analysis. 
Only deeper (and/or less contaminated) X-ray observations with \chandra\ or \xmm\ will allow us 
to properly characterise the diffuse X-ray emission and address this question.

\section{Summary} 
\label{sect. summary}

For the first time, we presented a study of the X-ray source populations 
and diffuse emission in the \xmm\ field of view of the galaxy pair NGC~1512 and NGC~1510.
We performed a source detection and obtained a catalogue containing 106 sources
detected in the energy range of $0.2-12$~keV, 15 within the $D_{25}$ ellipses
of NGC~1512 and NGC~1510, and about six sources in the extended arms of
the galaxy pair.
Based on the X-ray spectral analysis, Bayesian hardness ratios, X-ray variability
(for which we also made use of archival \sw/XRT data), cross-correlations with
catalogues in other wavelengths, and the 20-cm radio maps of ATCA,
we identified counterparts for 37 sources, six of which were identified or classified
as background objects and six as foreground stars.
We discussed the nature of source No.\,63 (XMMU~J040400.9-432319), previously classified as a QSO
\citep{Atlee07}. We have shown that its properties are also consistent with
an accreting black-hole in the galactic disc of NGC~1512. 

We showed that the number of HMXBs and LMXBs obtained from measurement
of the star formation rate and total mass of NGC~1512  
is consistent with the number of X-ray sources observed within the 
$D_{25}$ ellipse of NGC~1512, after taking into account the contribution
of background AGNs.

We detected diffuse X-ray emission from the interior region of \Gal\ 
with a plasma temperature $kT=0.68~(0.31-0.87)$~keV and a 0.3--10~keV X-ray luminosity 
of $1.3 \times 10^{38}$~erg~s$^{-1}$, after correcting for unresolved discrete sources. 
While the X-ray emission is most likely due to present and/or past stellar winds and supernovae, 
the derived electron densities and radiative cooling times seem to low and high, respectively, 
compared with those of other spiral galaxies.

\begin{acknowledgements}
This research is funded by the BMWi/DLR grant FKZ 50 OR 1209. 
MS and LD acknowledge support by the Deutsche Forschungsgemeinschaft
through the Emmy Noether Research Grant SA 2131/1-1.
This work is partially supported by the Bundesministerium für
Wirtschaft und Technologie through the Deutsches Zentrum für Luft-
und Raumfahrt (Grant FKZ 50 OG 1301).
This research has made use of the SIMBAD database,
operated at CDS, Strasbourg, France, and
of the NASA/IPAC Extragalactic Database (NED), 
which is operated by the Jet Propulsion Laboratory, 
California Institute of Technology, under contract 
with the National Aeronautics and Space Administration. 
This research has made use of the VizieR catalogue access tool, 
CDS, Strasbourg, France. The original description 
of the VizieR service was published in \citet{Ochsenbein00}.
This research has made use of data, software and web tools obtained 
from NASA's High Energy Astrophysics Science Archive Research Center (HEASARC), 
a service of Goddard Space Flight Center and the Smithsonian Astrophysical Observatory.
This publication has made use of data products from the 
Two Micron All Sky Survey, which is a joint project of the 
University of Massachusetts and the Infrared Processing and Analysis Center, 
funded by the National Aeronautics and 
Space Administration and the National Science Foundation.
This research has made use of SAOImage DS9, 
developed by Smithsonian Astrophysical Observatory. This work made use of the XMM-Newton Extended Source Analysis Software. 
\end{acknowledgements}

\bibliographystyle{aa} 
\bibliography{lducci_ngc1512}

\begin{appendix}

\section{classification and identification of the \emph{XMM-Newton} sources}
\label{sect. app class}

\subsection{foreground stars}
\label{sect. app. fgstars}

\paragraph{\textbf{Source No.\,9}}
Source No.\,9 is the brightest X-ray source in the 
\emph{XMM-Newton} observation of NGC 1512/1510.
It is located outside of the optical discs of NGC 1512 and NGC 1510 
(angular separation of $\sim 0.21^\circ$ from the nuclear region of NGC 1512).
It is associated with the optical source USNO-B1~0464-0031154
and the infrared source 2MASS~04030715-4330273.
It is close to the galaxy MRSS~250-130342 \citep{Ungruhe03} 
(angular separation of $15.3^{\prime\prime}$),
but the error circles do not overlap.
Therefore, the galaxy is probably not associated with source No.\,9.

Source No.\,9 is bright enough in the \emph{XMM-Newton} 
observation to allow a spectral analysis.
We fitted the PN, MOS1, MOS2 spectra simultaneously with the 
thermal palsma model of \citet{Kaastra00} (\texttt{mekal} in \texttt{XSPEC}),
which is commonly used to study the X-ray coronal emission from stars.
We first attempted to fit the spectrum with one-temperature model, but we obtained 
a poor fit with $\chi^2_\nu=1.25$.
Then, we fitted the spectrum with a two-temperature model, using non-solar abundances,
which led to a significant improvement of the fit (Table \ref{Tab. no.9 spectral parameters} and Fig. \ref{fig. no.9 23 54 63}).
We obtained the best-fit assuming metal abundances with regard to solar abundances of $0.2$.
Observations of under- or over-abundances of elements are not unusual 
in the coronae of stars (see e.g. \citealt{Mewe98}; \citealt{Favata98}).
We point out that in our fit the coronal abundances cannot be adequately constrained because of the poor statistics.

\begin{table}
\begin{center}
\caption{Best-fitting parameters of the X-ray spectra of source No.\,9 (errors at 90\% confidence level).
The flux (erg cm$^{-2}$ s$^{-1}$) has been calculated in the energy range $0.2-2$ keV.}
\label{Tab. no.9 spectral parameters}
\resizebox{\columnwidth}{!}{
\begin{tabular}{lcc}
\hline
\hline
\noalign{\smallskip}
                                 &      one-temperature model          &    two-temperature model         \\
\noalign{\smallskip}
\hline
\noalign{\smallskip}
$N_{\rm H}$ ($10^{22}$\,cm$^{-2}$)  &  $0.03{+0.02 \atop -0.02}$          &    $0.004{+0.010 \atop -0.004}$      \\
\noalign{\smallskip}
$kT_{\rm 1}$ (keV)                &  $0.84{+0.45 \atop -0.47}$           &     $0.61{+0.08 \atop -0.22}$        \\
\noalign{\smallskip}
norm.                           &  $5.1{+1.1 \atop -0.9}\times 10^{-4}$&$1.2{+0.4 \atop -0.5}\times 10^{-4}$   \\
\noalign{\smallskip}
$kT_{\rm 2}$ (keV)                &                                     &        $1.2{+0.2 \atop -0.2}$         \\
\noalign{\smallskip}
norm.                           &                                     &$2.2{+4.5 \atop -4.4}\times 10^{-4}$    \\
\noalign{\smallskip}
Abundances                      &  $0.09{+0.03 \atop -0.02}$          &                 $0.2$                  \\
\noalign{\smallskip}
$\chi^2_\nu$ (d.o.f.)           &          1.25 (81)                  &              1.04 (80)                \\
\noalign{\smallskip}
$F_{\rm x}$                      &$2.5{+1.2 \atop -0.8}\times 10^{-13}$&$2.6{+0.8 \atop -1.0}\times 10^{-13}$   \\
\noalign{\smallskip}
unabs. $F_{\rm x}$               &$3.4{+1.1 \atop -0.8}\times 10^{-13}$ &$2.9{+0.8 \atop -0.9}\times 10^{-13}$    \\
\noalign{\smallskip}
\hline
\end{tabular}
}
\end{center}
\end{table}

\paragraph{\textbf{Sources No.\,13 and 94}}
Sources No.\,13 and 94 have optical, infrared, and radio counterparts.
The radio counterparts, detected with ATCA, have fluxes of 0.9~mJy~beam$^{-1}$ (No.\,13)
and 0.3~mJy~beam$^{-1}$ (No.\,94) at 20-cm (beam size: $7.599\times5.470$~arcsec$^2$).
Although the measurement of a proper motion classify them as foreground stars\footnote{We point out 
that the measurement of the proper motion of No.\,94 in PPMXL catalogue ruled out 
the previous classification of this source as a galaxy 
(MRSS 250-123273 in the MRSS catalogue; APMUKS(BJ) B040305.24-433329.9 in the \emph{The APM galaxy survey}, \citealt{Maddox90}).}
they violate the $\log_{10}(f_{\rm x}/f_{\rm opt}) \leq -1$ (No.\,13)
and $HR_3 \lesssim -0.4$ and $J-K_{\rm s}$ vs. $B-R$ of Fig. \ref{fig. b-r_j-k.ps} (No.\,94) criteria.
The high $f_{\rm x}/f_{\rm opt}$ ratio of source No.\,13 can be explained by the strong X-ray and radio emission
of a low-mass pre-main sequence star (e.g. \citealt{Getman08}; \citealt{Carkner97} and references therein).
The hard $HR_3$ of No.\,94 may indicate a binary nature (with an accreting neutron star in a propeller state).

\subsection{background objects}
\label{sect. app. bg-obj}

\paragraph{\textbf{Source No.\,23}}

Source No.\,23 is located outside of the $D_{25}$ ellipse of NGC 1512 
($6.3^\prime$ from the nuclear region of NGC 1512).
It has neither an optical nor an infrared counterpart.
A radio counterpart has been detected with ATCA with a flux of 0.3~mJy~beam$^{-1}$ at 20-cm (beam size: $8.000 \times8.000$~arcsec$^2$).
The \emph{XMM-Newton} spectrum can be well fitted with an absorbed power-law
(see Table \ref{Tab. no.23 54 63 spectral parameters} and Fig. \ref{fig. no.9 23 54 63}).
Therefore, we classified source No.\,23 as an AGN candidate.

\section{X-ray source catalogue of the \emph{XMM-Newton} observation of the NGC\,1512/1510 system}
\label{sect. catalogue-table}

\ 

\onecolumn
\scriptsize
\begin{longtable}{lccccccccc}
\caption{\emph{(Only in the electronic version)} NGC\,1512/1510 sources detected by \emph{XMM-Newton} 
}
\label{Tab. source list}\\

\hline

\hline
\noalign{\smallskip}
No. &   RA(2000)  &  DEC(2000)   &          pos. err.      &        rate           & likelihood &    HR1   &   HR2  &  HR3  &   HR4     \\
    &             &              &                         &      ($0.2-12$\,keV)  &            &          &                &           \\
\hline
\endfirsthead

\multicolumn{10}{r}{Continued from previous page}\\
\hline
\noalign{\smallskip}
No. &   RA(2000)  &  DEC(2000)   &          pos. err.      &        rate           & likelihood &    HR1   &    HR2   &   HR3   &    HR4   \\
    &             &              &                         &     ($0.2-12$\,keV)   &            &          &          &         &          \\
\hline

\endhead

\hline \multicolumn{10}{r}{{Continued on next page}} \\
\endfoot

\hline
\endlastfoot

\hline
\noalign{\smallskip}
{ } { } 1 & 04 02 42.04 & $-$43 16 38.7 & $ 1.27^{\prime\prime}$ & $ 0.0042 \pm 0.0008$ & $  35.49$ & $ 0.35{+0.41 \atop -0.30}$ & $ 0.31{+0.25 \atop -0.22}$ & $-0.51{+0.23 \atop -0.30}$ & $-0.39{+0.15 \atop -0.61}$  \\ 
\noalign{\smallskip}
{ } { } 2 & 04 02 55.40 & $-$43 22 02.3 & $ 1.81^{\prime\prime}$ & $ 0.0047 \pm 0.0013$ & $   8.44$ & $ 0.32{+0.61 \atop -0.25}$ & $-0.15{+0.44 \atop -0.53}$ & $ 0.211{+0.54 \atop -0.38}$ & $-0.48{+0.13 \atop -0.52}$  \\ 
\noalign{\smallskip}
{ } { } 3 & 04 02 59.32 & $-$43 19 18.3 & $ 1.22^{\prime\prime}$ & $ 0.0073 \pm 0.0014$ & $  44.16$ & $ 0.30{+0.24 \atop -0.22}$ & $-0.14{+0.22 \atop -0.23}$ & $-0.45{+0.26 \atop -0.36}$ & $-0.34{+0.17 \atop -0.66}$  \\ 
\noalign{\smallskip}
{ } { } 4 & 04 02 59.40 & $-$43 32 23.6 & $ 1.38^{\prime\prime}$ & $ 0.0114 \pm 0.0029$ & $  21.72$ & $ 0.59{+0.36 \atop -0.14}$ & $-0.32{+0.28 \atop -0.31}$ & $-0.40{+0.15 \atop -0.60}$ & $ 0.01{+0.56 \atop -0.58}$  \\ 
\noalign{\smallskip}
{ } { } 5 & 04 03 02.96 & $-$43 15 47.3 & $ 2.03^{\prime\prime}$ & $ 0.0037 \pm 0.0012$ & $   8.07$ & $ 0.21{+0.79 \atop -0.26}$ & $-0.21{+0.26 \atop -0.79}$ & $ 0.51{+0.49 \atop -0.11}$ & $-0.44{+0.13 \atop -0.56}$  \\ 
\noalign{\smallskip}
{ } { } 6 & 04 03 03.12 & $-$43 24 15.4 & $ 1.26^{\prime\prime}$ & $ 0.0049 \pm 0.0013$ & $  10.76$ & $ 0.32{+0.48 \atop -0.30}$ & $-0.48{+0.13 \atop -0.53}$ & $ 0.45{+0.55 \atop -0.13}$ & $-0.40{+0.20 \atop -0.54}$  \\ 
\noalign{\smallskip}
{ } { } 7 & 04 03 03.72 & $-$43 28 18.5 & $ 1.00^{\prime\prime}$ & $ 0.0092 \pm 0.0016$ & $  31.61$ & $ 0.63{+0.37 \atop -0.08}$ & $ 0.22{+0.35 \atop -0.27}$ & $-0.46{+0.22 \atop -0.41}$ & $-0.24{+0.24 \atop -0.72}$  \\ 
\noalign{\smallskip}
{ } { } 8 & 04 03 05.55 & $-$43 21 18.4 & $ 1.18^{\prime\prime}$ & $ 0.0043 \pm 0.0011$ & $  22.18$ & $ 0.41{+0.43 \atop -0.24}$ & $ 0.19{+0.32 \atop -0.26}$ & $-0.65{+0.08 \atop -0.35}$ & $-0.16{+0.27 \atop -0.84}$  \\ 
\noalign{\smallskip}
{ } { } 9 & 04 03 07.15 & $-$43 30 29.0 & $ 0.18^{\prime\prime}$ & $ 0.2757 \pm 0.0063$ & $5692.49$ & $ 0.40{+0.03 \atop -0.03}$ & $-0.33{+0.03 \atop -0.03}$ & $-0.78{+0.05 \atop -0.05}$ & $-0.77{+0.08 \atop -0.22}$  \\ 
\noalign{\smallskip}
   { } 10 & 04 03 08.30 & $-$43 31 33.9 & $ 1.27^{\prime\prime}$ & $ 0.0088 \pm 0.0024$ & $  11.37$ & $ 0.51{+0.49 \atop -0.12}$ & $-0.18{+0.41 \atop -0.49}$ & $ 0.29{+0.52 \atop -0.31}$ & $-0.49{+0.12 \atop -0.51}$  \\ 
\noalign{\smallskip}
   { } 11 & 04 03 09.95 & $-$43 15 56.3 & $ 0.93^{\prime\prime}$ & $ 0.0103 \pm 0.0014$ & $  76.62$ & $ 0.21{+0.45 \atop -0.36}$ & $ 0.55{+0.21 \atop -0.18}$ & $ 0.03{+0.17 \atop -0.15}$ & $-0.73{+0.11 \atop -0.22}$  \\ 
\noalign{\smallskip}
   { } 12 & 04 03 11.98 & $-$43 20 32.6 & $ 1.77^{\prime\prime}$ & $ 0.0040 \pm 0.0010$ & $  11.61$ & $-0.02{+0.58 \atop -0.67}$ & $ 0.16{+0.84 \atop -0.27}$ & $ 0.58{+0.39 \atop -0.13}$ & $-0.58{+0.15 \atop -0.38}$  \\ 
\noalign{\smallskip}
   { } 13 & 04 03 12.96 & $-$43 11 57.6 & $ 0.89^{\prime\prime}$ & $ 0.0098 \pm 0.0015$ & $  50.03$ & $ 0.70{+0.30 \atop -0.06}$ & $ 0.22{+0.25 \atop -0.24}$ & $-0.34{+0.27 \atop -0.27}$ & $-0.46{+0.14 \atop -0.52}$  \\ 
\noalign{\smallskip}
   { } 14 & 04 03 13.46 & $-$43 28 02.0 & $ 1.35^{\prime\prime}$ & $ 0.0070 \pm 0.0016$ & $  13.91$ & $ 0.34{+0.66 \atop -0.18}$ & $-0.36{+0.17 \atop -0.64}$ & $ 0.60{+0.40 \atop -0.09}$ & $-0.48{+0.17 \atop -0.48}$  \\ 
\noalign{\smallskip}
   { } 15 & 04 03 17.87 & $-$43 14 58.0 & $ 0.94^{\prime\prime}$ & $ 0.0102 \pm 0.0015$ & $  68.50$ & $ 0.54{+0.27 \atop -0.19}$ & $ 0.11{+0.21 \atop -0.19}$ & $-0.50{+0.23 \atop -0.27}$ & $-0.25{+0.32 \atop -0.56}$  \\ 
\noalign{\smallskip}
   { } 16 & 04 03 20.13 & $-$43 25 36.9 & $ 1.50^{\prime\prime}$ & $ 0.0063 \pm 0.0012$ & $  19.95$ & $ 0.53{+0.47 \atop -0.09}$ & $ 0.56{+0.28 \atop -0.20}$ & $-0.35{+0.26 \atop -0.26}$ & $-0.11{+0.36 \atop -0.40}$  \\ 
\noalign{\smallskip}
   { } 17 & 04 03 20.51 & $-$43 17 59.9 & $ 0.84^{\prime\prime}$ & $ 0.0085 \pm 0.0010$ & $  80.36$ & $ 0.06{+0.18 \atop -0.19}$ & $-0.09{+0.19 \atop -0.19}$ & $-0.06{+0.21 \atop -0.21}$ & $-0.67{+0.08 \atop -0.33}$  \\ 
\noalign{\smallskip}
   { } 18 & 04 03 21.47 & $-$43 10 10.6 & $ 1.37^{\prime\prime}$ & $ 0.0057 \pm 0.0014$ & $  12.30$ & $ 0.46{+0.54 \atop -0.13}$ & $ 0.06{+0.33 \atop -0.36}$ & $-0.30{+0.34 \atop -0.42}$ & $-0.18{+0.36 \atop -0.63}$  \\ 
\noalign{\smallskip}
   { } 19 & 04 03 21.87 & $-$43 24 55.2 & $ 0.66^{\prime\prime}$ & $ 0.0199 \pm 0.0017$ & $ 149.95$ & $-0.09{+0.12 \atop -0.13}$ & $-0.29{+0.14 \atop -0.16}$ & $-0.51{+0.21 \atop -0.26}$ & $ 0.14{+0.37 \atop -0.36}$  \\ 
\noalign{\smallskip}
   { } 20 & 04 03 22.09 & $-$43 12 28.4 & $ 1.32^{\prime\prime}$ & $ 0.0036 \pm 0.0010$ & $   8.43$ & $ 0.42{+0.54 \atop -0.18}$ & $-0.02{+0.37 \atop -0.36}$ & $-0.55{+0.11 \atop -0.45}$ & $ 0.28{+0.72 \atop -0.20}$  \\ 
\noalign{\smallskip}
   { } 21 & 04 03 22.27 & $-$43 25 12.3 & $ 0.56^{\prime\prime}$ & $ 0.0349 \pm 0.0028$ & $ 294.29$ & $ 0.01{+0.16 \atop -0.16}$ & $-0.26{+0.19 \atop -0.19}$ & $ 0.13{+0.22 \atop -0.21}$ & $-0.81{+0.04 \atop -0.19}$  \\ 
\noalign{\smallskip}
   { } 22 & 04 03 23.19 & $-$43 17 17.3 & $ 0.88^{\prime\prime}$ & $ 0.0087 \pm 0.0011$ & $  63.40$ & $ 0.47{+0.39 \atop -0.23}$ & $ 0.10{+0.28 \atop -0.24}$ & $-0.29{+0.27 \atop -0.31}$ & $-0.09{+0.36 \atop -0.40}$  \\ 
\noalign{\smallskip}
   { } 23 & 04 03 25.37 & $-$43 17 22.0 & $ 0.30^{\prime\prime}$ & $ 0.0541 \pm 0.0020$ & $1709.88$ & $ 0.53{+0.05 \atop -0.05}$ & $-0.17{+0.05 \atop -0.05}$ & $-0.64{+0.06 \atop -0.06}$ & $-0.46{+0.18 \atop -0.17}$  \\ 
\noalign{\smallskip}
   { } 24 & 04 03 25.76 & $-$43 23 10.0 & $ 0.74^{\prime\prime}$ & $ 0.0112 \pm 0.0011$ & $ 136.76$ & $ 0.00{+0.17 \atop -0.17}$ & $-0.15{+0.19 \atop -0.19}$ & $-0.37{+0.29 \atop -0.28}$ & $-0.17{+0.39 \atop -0.50}$  \\ 
\noalign{\smallskip}
   { } 25 & 04 03 26.52 & $-$43 21 46.3 & $ 0.89^{\prime\prime}$ & $ 0.0049 \pm 0.0008$ & $  29.71$ & $ 0.15{+0.66 \atop -0.38}$ & $ 0.44{+0.39 \atop -0.25}$ & $-0.06{+0.31 \atop -0.29}$ & $-0.26{+0.34 \atop -0.41}$  \\ 
\noalign{\smallskip}
   { } 26 & 04 03 28.37 & $-$43 29 29.5 & $ 0.61^{\prime\prime}$ & $ 0.0220 \pm 0.0019$ & $ 203.27$ & $ 0.22{+0.13 \atop -0.13}$ & $-0.27{+0.13 \atop -0.13}$ & $-0.39{+0.23 \atop -0.21}$ & $-0.27{+0.30 \atop -0.38}$  \\ 
\noalign{\smallskip}
   { } 27 & 04 03 30.29 & $-$43 31 09.6 & $ 1.13^{\prime\prime}$ & $ 0.0092 \pm 0.0018$ & $  26.23$ & $ 0.42{+0.58 \atop -0.15}$ & $ 0.04{+0.42 \atop -0.41}$ & $ 0.07{+0.43 \atop -0.36}$ & $-0.49{+0.12 \atop -0.51}$  \\ 
\noalign{\smallskip}
   { } 28 & 04 03 32.12 & $-$43 21 39.2 & $ 0.38^{\prime\prime}$ & $ 0.0119 \pm 0.0010$ & $ 205.28$ & $ 0.80{+0.20 \atop -0.05}$ & $-0.10{+0.16 \atop -0.16}$ & $-0.28{+0.22 \atop -0.21}$ & $-0.67{+0.08 \atop -0.33}$  \\ 
\noalign{\smallskip}
   { } 29 & 04 03 32.71 & $-$43 24 01.7 & $ 1.30^{\prime\prime}$ & $ 0.0169 \pm 0.0015$ & $ 143.73$ & $ 0.39{+0.15 \atop -0.14}$ & $-0.60{+0.14 \atop -0.16}$ & $-0.10{+0.38 \atop -0.33}$ & $-0.23{+0.36 \atop -0.52}$  \\ 
\noalign{\smallskip}
   { } 30 & 04 03 35.40 & $-$43 17 45.8 & $ 0.48^{\prime\prime}$ & $ 0.0207 \pm 0.0013$ & $ 369.67$ & $ 0.12{+0.14 \atop -0.13}$ & $-0.26{+0.13 \atop -0.15}$ & $-0.47{+0.21 \atop -0.23}$ & $ 0.03{+0.38 \atop -0.36}$  \\ 
\noalign{\smallskip}
   { } 31 & 04 03 36.55 & $-$43 16 09.1 & $ 1.24^{\prime\prime}$ & $ 0.0047 \pm 0.0009$ & $  17.29$ & $ 0.25{+0.50 \atop -0.35}$ & $-0.28{+0.33 \atop -0.51}$ & $ 0.45{+0.43 \atop -0.22}$ & $-0.24{+0.33 \atop -0.37}$  \\ 
\noalign{\smallskip}
   { } 32 & 04 03 36.61 & $-$43 11 57.0 & $ 1.13^{\prime\prime}$ & $ 0.0070 \pm 0.0012$ & $  28.61$ & $ 0.13{+0.38 \atop -0.34}$ & $ 0.02{+0.35 \atop -0.34}$ & $-0.28{+0.34 \atop -0.45}$ & $-0.43{+0.14 \atop -0.58}$  \\ 
\noalign{\smallskip}
   { } 33 & 04 03 37.26 & $-$43 17 31.9 & $ 0.47^{\prime\prime}$ & $ 0.0129 \pm 0.0011$ & $ 209.76$ & $ 0.78{+0.22 \atop -0.05}$ & $ 0.13{+0.20 \atop -0.18}$ & $ 0.09{+0.17 \atop -0.17}$ & $-0.72{+0.09 \atop -0.27}$  \\ 
\noalign{\smallskip}
   { } 34 & 04 03 37.46 & $-$43 11 05.9 & $ 1.57^{\prime\prime}$ & $ 0.0043 \pm 0.0012$ & $   7.62$ & $-0.05{+0.33 \atop -0.31}$ & $-0.22{+0.32 \atop -0.40}$ & $-0.47{+0.13 \atop -0.53}$ & $ 0.33{+0.67 \atop -0.18}$  \\ 
\noalign{\smallskip}
   { } 35 & 04 03 37.63 & $-$43 27 38.8 & $ 0.94^{\prime\prime}$ & $ 0.0095 \pm 0.0013$ & $  60.41$ & $ 0.14{+0.26 \atop -0.26}$ & $ 0.10{+0.23 \atop -0.22}$ & $-0.32{+0.26 \atop -0.27}$ & $-0.29{+0.33 \atop -0.46}$  \\ 
\noalign{\smallskip}
   { } 36 & 04 03 37.69 & $-$43 09 09.0 & $ 1.10^{\prime\prime}$ & $ 0.0120 \pm 0.0017$ & $  57.48$ & $-0.11{+0.25 \atop -0.25}$ & $ 0.31{+0.21 \atop -0.21}$ & $-0.42{+0.23 \atop -0.23}$ & $-0.48{+0.15 \atop -0.49}$  \\ 
\noalign{\smallskip}
   { } 37 & 04 03 38.08 & $-$43 24 21.6 & $ 1.20^{\prime\prime}$ & $ 0.0026 \pm 0.0006$ & $   8.29$ & $ 0.22{+0.78 \atop -0.23}$ & $ 0.23{+0.53 \atop -0.34}$ & $-0.13{+0.40 \atop -0.49}$ & $-0.22{+0.30 \atop -0.69}$  \\ 
\noalign{\smallskip}
   { } 38 & 04 03 38.24 & $-$43 10 05.4 & $ 0.66^{\prime\prime}$ & $ 0.0199 \pm 0.0018$ & $ 168.78$ & $-0.04{+0.16 \atop -0.16}$ & $0.00{+0.17 \atop -0.18}$ & $-0.14{+0.20 \atop -0.19}$ & $-0.72{+0.07 \atop -0.28}$  \\ 
\noalign{\smallskip}
   { } 39 & 04 03 39.09 & $-$43 31 02.0 & $ 1.77^{\prime\prime}$ & $ 0.0073 \pm 0.0016$ & $  15.43$ & $ 0.34{+0.58 \atop -0.25}$ & $ 0.04{+0.41 \atop -0.37}$ & $-0.05{+0.37 \atop -0.42}$ & $-0.39{+0.19 \atop -0.58}$  \\ 
\noalign{\smallskip}
   { } 40 & 04 03 40.85 & $-$43 28 57.1 & $ 1.68^{\prime\prime}$ & $ 0.0082 \pm 0.0017$ & $  18.70$ & $ 0.29{+0.64 \atop -0.24}$ & $ 0.22{+0.44 \atop -0.34}$ & $-0.55{+0.11 \atop -0.45}$ & $ 0.01{+0.68 \atop -0.61}$  \\ 
\noalign{\smallskip}
   { } 41 & 04 03 41.82 & $-$43 10 18.3 & $ 0.71^{\prime\prime}$ & $ 0.0215 \pm 0.0019$ & $ 200.03$ & $ 0.29{+0.22 \atop -0.22}$ & $ 0.22{+0.16 \atop -0.15}$ & $-0.25{+0.16 \atop -0.16}$ & $-0.52{+0.20 \atop -0.24}$  \\ 
\noalign{\smallskip}
   { } 42 & 04 03 42.10 & $-$43 18 38.5 & $ 1.18^{\prime\prime}$ & $ 0.0029 \pm 0.0006$ & $  11.49$ & $ 0.33{+0.67 \atop -0.19}$ & $ 0.45{+0.49 \atop -0.19}$ & $-0.40{+0.24 \atop -0.50}$ & $ 0.06{+0.61 \atop -0.48}$  \\ 
\noalign{\smallskip}
   { } 43 & 04 03 42.40 & $-$43 19 58.9 & $ 0.55^{\prime\prime}$ & $ 0.0105 \pm 0.0009$ & $ 166.39$ & $ 0.26{+0.15 \atop -0.14}$ & $-0.32{+0.15 \atop -0.15}$ & $-0.69{+0.08 \atop -0.31}$ & $ 0.12{+0.72 \atop -0.39}$  \\ 
\noalign{\smallskip}
   { } 44 & 04 03 42.72 & $-$43 23 17.1 & $ 1.46^{\prime\prime}$ & $ 0.0025 \pm 0.0006$ & $  10.36$ & $-0.04{+0.53 \atop -0.62}$ & $ 0.08{+0.68 \atop -0.46}$ & $ 0.44{+0.49 \atop -0.20}$ & $-0.56{+0.11 \atop -0.44}$  \\ 
\noalign{\smallskip}
   { } 45 & 04 03 42.90 & $-$43 20 50.2 & $ 1.41^{\prime\prime}$ & $ 0.0027 \pm 0.0006$ & $   8.80$ & $ 0.44{+0.57 \atop -0.14}$ & $ 0.01{+0.36 \atop -0.39}$ & $ 0.06{+0.40 \atop -0.33}$ & $-0.39{+0.18 \atop -0.59}$  \\ 
\noalign{\smallskip}
   { } 46 & 04 03 42.94 & $-$43 16 01.3 & $ 0.38^{\prime\prime}$ & $ 0.0148 \pm 0.0011$ & $ 272.46$ & $ 0.13{+0.14 \atop -0.14}$ & $-0.08{+0.13 \atop -0.13}$ & $-0.27{+0.16 \atop -0.16}$ & $-0.55{+0.21 \atop -0.31}$  \\ 
\noalign{\smallskip}
   { } 47 & 04 03 45.43 & $-$43 26 31.9 & $ 1.14^{\prime\prime}$ & $ 0.0062 \pm 0.0013$ & $  20.24$ & $ 0.48{+0.52 \atop -0.13}$ & $ 0.18{+0.47 \atop -0.36}$ & $-0.26{+0.34 \atop -0.51}$ & $-0.25{+0.21 \atop -0.75}$  \\ 
\noalign{\smallskip}
   { } 48 & 04 03 47.11 & $-$43 24 42.8 & $ 1.52^{\prime\prime}$ & $ 0.0032 \pm 0.0008$ & $   6.59$ & $-0.19{+0.27 \atop -0.81}$ & $ 0.28{+0.72 \atop -0.22}$ & $ 0.24{+0.69 \atop -0.27}$ & $ 0.02{+0.49 \atop -0.47}$  \\ 
\noalign{\smallskip}
   { } 49 & 04 03 48.29 & $-$43 10 11.7 & $ 1.20^{\prime\prime}$ & $ 0.0092 \pm 0.0014$ & $  37.38$ & $ 0.47{+0.53 \atop -0.13}$ & $ 0.38{+0.26 \atop -0.25}$ & $-0.52{+0.24 \atop -0.31}$ & $-0.33{+0.17 \atop -0.67}$  \\ 
\noalign{\smallskip}
   { } 50 & 04 03 48.97 & $-$43 14 54.4 & $ 0.78^{\prime\prime}$ & $ 0.0068 \pm 0.0009$ & $  48.67$ & $-0.15{+0.37 \atop -0.44}$ & $ 0.31{+0.42 \atop -0.28}$ & $ 0.19{+0.30 \atop -0.29}$ & $-0.52{+0.18 \atop -0.41}$  \\ 
\noalign{\smallskip}
   { } 51 & 04 03 50.86 & $-$43 22 04.9 & $ 1.54^{\prime\prime}$ & $ 0.0027 \pm 0.0006$ & $  13.36$ & $-0.09{+0.45 \atop -0.70}$ & $ 0.55{+0.45 \atop -0.11}$ & $-0.38{+0.29 \atop -0.41}$ & $-0.18{+0.32 \atop -0.73}$  \\ 
\noalign{\smallskip}
   { } 52 & 04 03 50.91 & $-$43 24 40.1 & $ 1.07^{\prime\prime}$ & $ 0.0031 \pm 0.0007$ & $  14.07$ & $ 0.13{+0.87 \atop -0.29}$ & $ 0.38{+0.62 \atop -0.16}$ & $ 0.29{+0.37 \atop -0.31}$ & $-0.52{+0.13 \atop -0.48}$  \\ 
\noalign{\smallskip}
   { } 53 & 04 03 52.76 & $-$43 20 59.0 & $ 1.31^{\prime\prime}$ & $ 0.0030 \pm 0.0009$ & $   7.72$ & $ 0.14{+0.42 \atop -0.37}$ & $-0.27{+0.32 \atop -0.47}$ & $-0.03{+0.50 \atop -0.52}$ & $-0.22{+0.23 \atop -0.78}$  \\ 
\noalign{\smallskip}
   { } 54 & 04 03 54.26 & $-$43 20 56.7 & $ 0.21^{\prime\prime}$ & $ 0.0697 \pm 0.0020$ & $2705.45$ & $ 0.22{+0.06 \atop -0.05}$ & $-0.31{+0.05 \atop -0.05}$ & $-0.37{+0.07 \atop -0.08}$ & $-0.46{+0.14 \atop -0.13}$  \\ 
\noalign{\smallskip}
   { } 55 & 04 03 54.30 & $-$43 18 53.0 & $ 0.94^{\prime\prime}$ & $ 0.0044 \pm 0.0007$ & $  29.10$ & $ 0.33{+0.50 \atop -0.30}$ & $ 0.24{+0.31 \atop -0.28}$ & $-0.14{+0.28 \atop -0.29}$ & $-0.22{+0.37 \atop -0.39}$  \\ 
\noalign{\smallskip}
   { } 56 & 04 03 54.52 & $-$43 14 32.6 & $ 1.76^{\prime\prime}$ & $ 0.0031 \pm 0.0007$ & $  10.63$ & $-0.43{+0.14 \atop -0.57}$ & $ 0.19{+0.81 \atop -0.25}$ & $ 0.53{+0.47 \atop -0.12}$ & $-0.30{+0.33 \atop -0.40}$  \\ 
\noalign{\smallskip}
   { } 57 & 04 03 54.79 & $-$43 22 24.5 & $ 1.42^{\prime\prime}$ & $ 0.0048 \pm 0.0019$ & $   6.25$ & $ 0.34{+0.66 \atop -0.19}$ & $ 0.26{+0.57 \atop -0.32}$ & $-0.34{+0.18 \atop -0.66}$ & $ 0.15{+0.80 \atop -0.30}$  \\ 
\noalign{\smallskip}
   { } 58 & 04 03 55.58 & $-$43 09 45.8 & $ 0.44^{\prime\prime}$ & $ 0.0458 \pm 0.0025$ & $ 706.75$ & $ 0.07{+0.08 \atop -0.08}$ & $-0.30{+0.09 \atop -0.09}$ & $-0.51{+0.15 \atop -0.16}$ & $-0.18{+0.26 \atop -0.29}$  \\ 
\noalign{\smallskip}
   { } 59 & 04 03 56.14 & $-$43 17 20.7 & $ 0.52^{\prime\prime}$ & $ 0.0143 \pm 0.0010$ & $ 286.01$ & $ 0.26{+0.12 \atop -0.12}$ & $-0.29{+0.11 \atop -0.12}$ & $-0.32{+0.18 \atop -0.18}$ & $-0.46{+0.27 \atop -0.37}$  \\ 
\noalign{\smallskip}
   { } 60 & 04 03 56.36 & $-$43 08 30.7 & $ 1.08^{\prime\prime}$ & $ 0.0168 \pm 0.0037$ & $  37.73$ & $ 0.56{+0.44 \atop -0.11}$ & $ 0.25{+0.32 \atop -0.26}$ & $ 0.19{+0.24 \atop -0.23}$ & $-0.52{+0.20 \atop -0.24}$  \\ 
\noalign{\smallskip}
   { } 61 & 04 03 56.60 & $-$43 29 05.0 & $ 1.61^{\prime\prime}$ & $ 0.0056 \pm 0.0011$ & $  16.82$ & $ 0.25{+0.33 \atop -0.31}$ & $-0.65{+0.08 \atop -0.35}$ & $ 0.14{+0.82 \atop -0.31}$ & $-0.30{+0.19 \atop -0.70}$  \\ 
\noalign{\smallskip}
   { } 62 & 04 03 58.34 & $-$43 22 46.0 & $ 1.20^{\prime\prime}$ & $ 0.0047 \pm 0.0008$ & $  26.37$ & $ 0.19{+0.52 \atop -0.38}$ & $ 0.20{+0.35 \atop -0.35}$ & $ 0.47{+0.19 \atop -0.17}$ & $-0.61{+0.19 \atop -0.24}$  \\ 
\noalign{\smallskip}
   { } 63 & 04 04 00.99 & $-$43 23 19.0 & $ 0.23^{\prime\prime}$ & $ 0.0687 \pm 0.0023$ & $2309.37$ & $-0.01{+0.06 \atop -0.05}$ & $-0.23{+0.06 \atop -0.06}$ & $-0.52{+0.08 \atop -0.09}$ & $-0.49{+0.20 \atop -0.19}$  \\ 
\noalign{\smallskip}
   { } 64 & 04 04 01.23 & $-$43 33 23.2 & $ 1.98^{\prime\prime}$ & $ 0.0211 \pm 0.0092$ & $  12.21$ & $-0.13{+0.37 \atop -0.49}$ & $ 0.21{+0.47 \atop -0.33}$ & $-0.41{+0.15 \atop -0.59}$ & $-0.39{+0.15 \atop -0.61}$  \\ 
\noalign{\smallskip}
   { } 65 & 04 04 03.38 & $-$43 25 38.4 & $ 1.20^{\prime\prime}$ & $ 0.0043 \pm 0.0009$ & $  19.54$ & $ 0.26{+0.74 \atop -0.21}$ & $ 0.36{+0.48 \atop -0.28}$ & $ 0.06{+0.35 \atop -0.33}$ & $-0.37{+0.28 \atop -0.45}$  \\ 
\noalign{\smallskip}
   { } 66 & 04 04 08.28 & $-$43 14 14.9 & $ 1.04^{\prime\prime}$ & $ 0.0075 \pm 0.0025$ & $  27.53$ & $-0.14{+0.40 \atop -0.53}$ & $ 0.63{+0.29 \atop -0.15}$ & $-0.59{+0.17 \atop -0.34}$ & $-0.34{+0.18 \atop -0.66}$  \\ 
\noalign{\smallskip}
   { } 67 & 04 04 08.90 & $-$43 18 43.8 & $ 1.31^{\prime\prime}$ & $ 0.0027 \pm 0.0006$ & $   7.70$ & $ 0.28{+0.37 \atop -0.30}$ & $-0.47{+0.22 \atop -0.42}$ & $ 0.19{+0.55 \atop -0.38}$ & $-0.43{+0.15 \atop -0.57}$  \\ 
\noalign{\smallskip}
   { } 68 & 04 04 10.89 & $-$43 07 03.6 & $ 2.10^{\prime\prime}$ & $ 0.0170 \pm 0.0025$ & $  50.21$ & $ 0.30{+0.70 \atop -0.20}$ & $ 0.55{+0.35 \atop -0.18}$ & $ 0.01{+0.25 \atop -0.26}$ & $-0.56{+0.19 \atop -0.31}$  \\ 
\noalign{\smallskip}
   { } 69 & 04 04 11.17 & $-$43 19 57.7 & $ 0.72^{\prime\prime}$ & $ 0.0083 \pm 0.0009$ & $  95.23$ & $ 0.31{+0.26 \atop -0.25}$ & $ 0.24{+0.18 \atop -0.18}$ & $-0.40{+0.20 \atop -0.19}$ & $-0.50{+0.13 \atop -0.50}$  \\ 
\noalign{\smallskip}
   { } 70 & 04 04 11.19 & $-$43 16 58.8 & $ 0.82^{\prime\prime}$ & $ 0.0068 \pm 0.0009$ & $  66.41$ & $ 0.47{+0.28 \atop -0.23}$ & $-0.05{+0.21 \atop -0.22}$ & $-0.28{+0.28 \atop -0.27}$ & $-0.32{+0.31 \atop -0.48}$  \\ 
\noalign{\smallskip}
   { } 71 & 04 04 12.64 & $-$43 22 30.7 & $ 0.81^{\prime\prime}$ & $ 0.0068 \pm 0.0008$ & $  79.81$ & $ 0.70{+0.30 \atop -0.07}$ & $ 0.03{+0.22 \atop -0.21}$ & $-0.12{+0.22 \atop -0.24}$ & $-0.42{+0.30 \atop -0.40}$  \\ 
\noalign{\smallskip}
   { } 72 & 04 04 13.90 & $-$43 29 00.1 & $ 1.13^{\prime\prime}$ & $ 0.0121 \pm 0.0017$ & $  67.72$ & $ 0.79{+0.21 \atop -0.04}$ & $ 0.01{+0.24 \atop -0.23}$ & $-0.03{+0.24 \atop -0.25}$ & $-0.76{+0.05 \atop -0.24}$  \\ 
\noalign{\smallskip}
   { } 73 & 04 04 14.97 & $-$43 22 57.6 & $ 0.79^{\prime\prime}$ & $ 0.0083 \pm 0.0010$ & $  77.13$ & $-0.05{+0.21 \atop -0.21}$ & $-0.29{+0.24 \atop -0.26}$ & $-0.01{+0.35 \atop -0.33}$ & $-0.40{+0.23 \atop -0.51}$  \\ 
\noalign{\smallskip}
   { } 74 & 04 04 15.17 & $-$43 22 34.3 & $ 1.61^{\prime\prime}$ & $ 0.0025 \pm 0.0007$ & $   9.05$ & $ 0.60{+0.40 \atop -0.10}$ & $ 0.17{+0.31 \atop -0.28}$ & $-0.51{+0.19 \atop -0.39}$ & $ 0.14{+0.60 \atop -0.39}$  \\ 
\noalign{\smallskip}
   { } 75 & 04 04 15.21 & $-$43 24 09.8 & $ 1.40^{\prime\prime}$ & $ 0.0028 \pm 0.0008$ & $   8.97$ & $ 0.33{+0.38 \atop -0.28}$ & $-0.19{+0.33 \atop -0.35}$ & $-0.42{+0.15 \atop -0.58}$ & $ 0.08{+0.77 \atop -0.42}$  \\ 
\noalign{\smallskip}
   { } 76 & 04 04 15.72 & $-$43 31 49.2 & $ 1.23^{\prime\prime}$ & $ 0.0109 \pm 0.0021$ & $  33.95$ & $ 0.27{+0.33 \atop -0.32}$ & $-0.41{+0.28 \atop -0.37}$ & $-0.22{+0.27 \atop -0.73}$ & $-0.22{+0.23 \atop -0.78}$  \\ 
\noalign{\smallskip}
   { } 77 & 04 04 15.79 & $-$43 28 03.7 & $ 0.87^{\prime\prime}$ & $ 0.0120 \pm 0.0015$ & $  77.02$ & $ 0.30{+0.30 \atop -0.25}$ & $ 0.13{+0.24 \atop -0.22}$ & $-0.46{+0.21 \atop -0.28}$ & $-0.28{+0.33 \atop -0.55}$  \\ 
\noalign{\smallskip}
   { } 78 & 04 04 17.53 & $-$43 19 06.8 & $ 1.18^{\prime\prime}$ & $ 0.0033 \pm 0.0007$ & $  10.36$ & $ 0.29{+0.65 \atop -0.24}$ & $-0.15{+0.39 \atop -0.55}$ & $-0.19{+0.30 \atop -0.73}$ & $ 0.06{+0.69 \atop -0.49}$  \\ 
\noalign{\smallskip}
   { } 79 & 04 04 17.77 & $-$43 25 47.5 & $ 0.95^{\prime\prime}$ & $ 0.0108 \pm 0.0014$ & $  65.43$ & $ 0.66{+0.34 \atop -0.09}$ & $ 0.34{+0.19 \atop -0.18}$ & $-0.39{+0.18 \atop -0.19}$ & $-0.36{+0.30 \atop -0.36}$  \\ 
\noalign{\smallskip}
   { } 80 & 04 04 17.94 & $-$43 29 13.2 & $ 1.18^{\prime\prime}$ & $ 0.0070 \pm 0.0016$ & $  19.29$ & $ 0.00{+0.57 \atop -0.50}$ & $ 0.44{+0.46 \atop -0.22}$ & $ 0.10{+0.29 \atop -0.29}$ & $-0.65{+0.08 \atop -0.35}$  \\ 
\noalign{\smallskip}
   { } 81 & 04 04 20.21 & $-$43 24 30.3 & $ 1.04^{\prime\prime}$ & $ 0.0048 \pm 0.0009$ & $  19.93$ & $ 0.17{+0.71 \atop -0.34}$ & $ 0.45{+0.42 \atop -0.22}$ & $ 0.07{+0.30 \atop -0.29}$ & $-0.64{+0.09 \atop -0.36}$  \\ 
\noalign{\smallskip}
   { } 82 & 04 04 21.24 & $-$43 22 23.4 & $ 1.62^{\prime\prime}$ & $ 0.0029 \pm 0.0008$ & $   7.20$ & $ 0.30{+0.66 \atop -0.22}$ & $-0.36{+0.17 \atop -0.64}$ & $ 0.36{+0.65 \atop -0.17}$ & $-0.32{+0.18 \atop -0.68}$  \\ 
\noalign{\smallskip}
   { } 83 & 04 04 22.57 & $-$43 24 52.3 & $ 1.70^{\prime\prime}$ & $ 0.0036 \pm 0.0009$ & $  10.77$ & $-0.03{+0.57 \atop -0.66}$ & $ 0.50{+0.50 \atop -0.12}$ & $ 0.12{+0.36 \atop -0.32}$ & $-0.63{+0.09 \atop -0.37}$  \\ 
\noalign{\smallskip}
   { } 84 & 04 04 23.05 & $-$43 12 02.0 & $ 1.18^{\prime\prime}$ & $ 0.0063 \pm 0.0012$ & $  22.66$ & $ 0.12{+0.36 \atop -0.34}$ & $-0.17{+0.35 \atop -0.38}$ & $ 0.09{+0.39 \atop -0.38}$ & $-0.48{+0.13 \atop -0.52}$  \\ 
\noalign{\smallskip}
   { } 85 & 04 04 24.15 & $-$43 25 46.3 & $ 1.42^{\prime\prime}$ & $ 0.0045 \pm 0.0012$ & $   8.05$ & $ 0.47{+0.44 \atop -0.20}$ & $-0.24{+0.35 \atop -0.38}$ & $-0.17{+0.38 \atop -0.56}$ & $-0.28{+0.20 \atop -0.73}$  \\ 
\noalign{\smallskip}
   { } 86 & 04 04 24.20 & $-$43 13 43.2 & $ 1.21^{\prime\prime}$ & $ 0.0072 \pm 0.0014$ & $  19.86$ & $-0.05{+0.34 \atop -0.36}$ & $-0.11{+0.37 \atop -0.40}$ & $-0.24{+0.34 \atop -0.58}$ & $ 0.32{+0.54 \atop -0.29}$  \\ 
\noalign{\smallskip}
   { } 87 & 04 04 26.82 & $-$43 28 40.7 & $ 1.03^{\prime\prime}$ & $ 0.0077 \pm 0.0015$ & $  25.75$ & $ 0.29{+0.29 \atop -0.25}$ & $-0.34{+0.28 \atop -0.31}$ & $-0.45{+0.14 \atop -0.56}$ & $-0.13{+0.28 \atop -0.87}$  \\ 
\noalign{\smallskip}
   { } 88 & 04 04 29.73 & $-$43 23 13.9 & $ 1.43^{\prime\prime}$ & $ 0.0030 \pm 0.0009$ & $   7.51$ & $ 0.22{+0.49 \atop -0.37}$ & $-0.11{+0.42 \atop -0.43}$ & $-0.20{+0.35 \atop -0.64}$ & $-0.16{+0.26 \atop -0.84}$  \\ 
\noalign{\smallskip}
   { } 89 & 04 04 31.49 & $-$43 26 10.4 & $ 1.42^{\prime\prime}$ & $ 0.0072 \pm 0.0018$ & $  18.81$ & $ 0.11{+0.37 \atop -0.34}$ & $-0.58{+0.10 \atop -0.42}$ & $ 0.03{+0.77 \atop -0.50}$ & $ 0.09{+0.77 \atop -0.43}$  \\ 
\noalign{\smallskip}
   { } 90 & 04 04 35.29 & $-$43 20 06.7 & $ 1.33^{\prime\prime}$ & $ 0.0054 \pm 0.0011$ & $  22.06$ & $ 0.39{+0.56 \atop -0.19}$ & $ 0.44{+0.26 \atop -0.24}$ & $-0.57{+0.20 \atop -0.30}$ & $-0.32{+0.18 \atop -0.68}$  \\ 
\noalign{\smallskip}
   { } 91 & 04 04 37.22 & $-$43 13 22.1 & $ 0.99^{\prime\prime}$ & $ 0.0079 \pm 0.0013$ & $  30.85$ & $ 0.31{+0.34 \atop -0.30}$ & $-0.20{+0.29 \atop -0.31}$ & $-0.30{+0.33 \atop -0.47}$ & $-0.42{+0.14 \atop -0.58}$  \\ 
\noalign{\smallskip}
   { } 92 & 04 04 38.05 & $-$43 14 03.7 & $ 0.44^{\prime\prime}$ & $ 0.0388 \pm 0.0023$ & $ 510.49$ & $ 0.21{+0.10 \atop -0.09}$ & $-0.17{+0.09 \atop -0.10}$ & $-0.54{+0.15 \atop -0.14}$ & $-0.41{+0.26 \atop -0.34}$  \\ 
\noalign{\smallskip}
   { } 93 & 04 04 41.89 & $-$43 09 43.4 & $ 0.78^{\prime\prime}$ & $ 0.0122 \pm 0.0014$ & $ 162.33$ & $ 0.18{+0.22 \atop -0.20}$ & $ 0.19{+0.18 \atop -0.16}$ & $-0.40{+0.17 \atop -0.20}$ & $-0.67{+0.08 \atop -0.33}$  \\ 
\noalign{\smallskip}
   { } 94 & 04 04 43.73 & $-$43 25 22.9 & $ 1.43^{\prime\prime}$ & $ 0.0057 \pm 0.0015$ & $  12.83$ & $-0.55{+0.11 \atop -0.45}$ & $ 0.18{+0.83 \atop -0.28}$ & $ 0.17{+0.79 \atop -0.28}$ & $-0.17{+0.35 \atop -0.70}$  \\ 
\noalign{\smallskip}
   { } 95 & 04 04 46.15 & $-$43 15 37.8 & $ 1.77^{\prime\prime}$ & $ 0.0052 \pm 0.0012$ & $  11.55$ & $ 0.47{+0.54 \atop -0.13}$ & $ 0.20{+0.40 \atop -0.34}$ & $-0.26{+0.34 \atop -0.42}$ & $-0.19{+0.36 \atop -0.57}$  \\ 
\noalign{\smallskip}
   { } 96 & 04 04 46.86 & $-$43 29 11.3 & $ 1.06^{\prime\prime}$ & $ 0.0089 \pm 0.0020$ & $  16.51$ & $ 0.47{+0.53 \atop -0.13}$ & $-0.14{+0.40 \atop -0.46}$ & $ 0.05{+0.49 \atop -0.44}$ & $-0.50{+0.11 \atop -0.50}$  \\ 
\noalign{\smallskip}
   { } 97 & 04 04 47.02 & $-$43 16 30.6 & $ 1.32^{\prime\prime}$ & $ 0.0084 \pm 0.0013$ & $  30.08$ & $-0.16{+0.27 \atop -0.27}$ & $ 0.09{+0.28 \atop -0.30}$ & $-0.33{+0.33 \atop -0.35}$ & $-0.44{+0.13 \atop -0.56}$  \\ 
\noalign{\smallskip}
   { } 98 & 04 04 47.14 & $-$43 22 03.5 & $ 0.88^{\prime\prime}$ & $ 0.0126 \pm 0.0017$ & $  61.83$ & $-0.56{+0.10 \atop -0.44}$ & $ 0.40{+0.60 \atop -0.15}$ & $ 0.46{+0.37 \atop -0.23}$ & $-0.25{+0.27 \atop -0.31}$  \\ 
\noalign{\smallskip}
   { } 99 & 04 04 49.77 & $-$43 20 43.8 & $ 0.88^{\prime\prime}$ & $ 0.0150 \pm 0.0018$ & $  84.92$ & $-0.01{+0.29 \atop -0.26}$ & $ 0.37{+0.20 \atop -0.20}$ & $-0.23{+0.21 \atop -0.19}$ & $-0.49{+0.23 \atop -0.31}$  \\ 
\noalign{\smallskip}
      100 & 04 04 51.53 & $-$43 19 03.8 & $ 1.32^{\prime\prime}$ & $ 0.0088 \pm 0.0016$ & $  24.31$ & $-0.08{+0.29 \atop -0.29}$ & $-0.05{+0.32 \atop -0.33}$ & $ 0.07{+0.32 \atop -0.31}$ & $-0.45{+0.24 \atop -0.40}$  \\ 
\noalign{\smallskip}
      101 & 04 04 59.35 & $-$43 13 32.7 & $ 0.69^{\prime\prime}$ & $ 0.0091 \pm 0.0010$ & $ 150.06$ & $ 0.40{+0.13 \atop -0.13}$ & $-0.25{+0.13 \atop -0.13}$ & $-0.22{+0.18 \atop -0.18}$ & $-0.56{+0.20 \atop -0.27}$  \\ 
\noalign{\smallskip}
      102 & 04 05 00.38 & $-$43 14 46.5 & $ 1.06^{\prime\prime}$ & $ 0.0108 \pm 0.0021$ & $  34.43$ & $-0.11{+0.46 \atop -0.52}$ & $ 0.40{+0.44 \atop -0.26}$ & $ 0.05{+0.30 \atop -0.30}$ & $-0.64{+0.08 \atop -0.36}$  \\ 
\noalign{\smallskip}
      103 & 04 05 02.17 & $-$43 20 36.9 & $ 2.00^{\prime\prime}$ & $ 0.0082 \pm 0.0019$ & $   7.55$ & $ 0.17{+0.59 \atop -0.38}$ & $ 0.06{+0.48 \atop -0.43}$ & $-0.16{+0.42 \atop -0.55}$ & $-0.03{+0.44 \atop -0.56}$  \\ 
\noalign{\smallskip}
      104 & 04 05 07.07 & $-$43 18 31.6 & $ 1.26^{\prime\prime}$ & $ 0.0088 \pm 0.0019$ & $  24.88$ & $ 0.29{+0.29 \atop -0.26}$ & $-0.61{+0.11 \atop -0.38}$ & $ 0.11{+0.65 \atop -0.43}$ & $-0.38{+0.15 \atop -0.63}$  \\ 
\noalign{\smallskip}
      105 & 04 05 09.57 & $-$43 18 50.7 & $ 0.95^{\prime\prime}$ & $ 0.0258 \pm 0.0028$ & $ 150.49$ & $ 0.31{+0.14 \atop -0.14}$ & $-0.25{+0.14 \atop -0.13}$ & $-0.60{+0.20 \atop -0.24}$ & $-0.23{+0.32 \atop -0.56}$  \\ 
\noalign{\smallskip}
      106 & 04 05 11.59 & $-$43 22 34.0 & $ 2.19^{\prime\prime}$ & $ 0.0086 \pm 0.0021$ & $  15.21$ & $-0.32{+0.28 \atop -0.32}$ & $ 0.18{+0.38 \atop -0.33}$ & $-0.60{+0.10 \atop -0.40}$ & $-0.25{+0.24 \atop -0.76}$  \\ 
\noalign{\smallskip}
\hline
\end{longtable}

\scriptsize
\begin{longtable}{lccccc}
\caption{\emph{(Only in the electronic version)} NGC\,1512/1510 sources detected by \emph{XMM-Newton} cross-correlated with 
optical, infrared, and radio counterparts.
For each source the identification proposed by the respective authors and our classification are given.
Uncertain classifications are given in brackets. }
\label{Tab. source list classification}\\

\hline
\multicolumn{1}{l}{No.} &
\multicolumn{1}{c}{USNO B1} &
\multicolumn{1}{c}{2MASS} &
\multicolumn{1}{c}{optical} &
\multicolumn{1}{c}{radio} &
\multicolumn{1}{c}{class.}\\
\hline
\endfirsthead

\multicolumn{6}{r}{Continued from previous page}\\
\hline
\multicolumn{1}{l}{No.} &
\multicolumn{1}{c}{USNO B1} &
\multicolumn{1}{c}{2MASS} &
\multicolumn{1}{c}{optical} &
\multicolumn{1}{c}{radio} &
\multicolumn{1}{c}{class.}\\
\hline

\endhead

\hline \multicolumn{6}{r}{{Continued on next page}} \\
\endfoot

\hline
\endlastfoot
{ } { } 1 & 0467-0032009 &                  &                        &                              &     \\
\hline
{ } { } 2 &              &                  &                        &                              &      \\
\hline
{ } { } 3 &              &                  &                        &                              &       \\
\hline
{ } { } 4 & 0464-0031137 & 04025923-4332221 &                        &                              &  foreground star   \\
\hline
{ } { } 5 &              &                  &                        &                              &    \\
\hline
{ } { } 6 &              &                  &                        &                              &     \\ 
\hline
{ } { } 7 &              &                  &                        &                              &     \\
\hline
{ } { } 8 &              &                  &                        &                              &    \\
\hline
{ } { } 9 & 0464-0031154 & 04030715-4330273 &                        &                              & foreground star   \\
\hline
{ } 10    & 0464-0031159 &                  &                        &                              &     \\
\hline
{ } 11    &              &                  &                        &                              &    \\
\hline
{ } 12    &              &                  &                        &                              &    \\
\hline
{ } 13    & 0468-0031995 & 04031316-4311583 &                        & ATCA 04 03 12.92 -43 11 56.0 & foreground star   \\
\hline
{ } 14    &              &                  &                        &                              &     \\
\hline
{ } 15    &              &                  &                        &                              &    \\
\hline
{ } 16    &              &                  &                        &                              &    \\
\hline
{ } 17    &              &                  &                        &                              &    \\
\hline
{ } 18    &              &                  &                        &                              &     \\
\hline
{ } 19    & 0465-0031058 &                  &                        &                              &       \\ 
\hline
{ } 20    & 0467-0032091 &                  &                        &                              &    \\
\hline
{ } 21    &              &                  &                        &                              &    \\
\hline
{ } 22    &              &                  &                        &                              &    \\
\hline
{ } 23    &              &                  &                        & ATCA 04 03 25.54 -43 17 23.2 & background object   \\
\hline
{ } 24    &              &                  &                        &                              &    \\
\hline
{ } 25    &              &                  &                        &                              &   \\
\hline
{ } 26    & 0465-0031069 &                  &                        &                              &    \\
\hline
{ } 27    &              &                  &                        &                              &    \\
\hline
{ } 28    &              &                  &                        &                              &    \\
\hline
{ } 29    &              &                  &      NGC1510           & ATCA 04 03 32.69 -43 23 59.2 & NGC 1510     \\
\hline
{ } 30    &              &                  &                        &                              &     \\
\hline
{ } 31    &              &                  &                        &                              &    \\
\hline
{ } 32    &              &                  &                        &                              &    \\
\hline
{ } 33    &              &                  &                        &                              &    \\
\hline
{ } 34    &              &                  &                        &                              &    \\
\hline
{ } 35    &              &                  &                        &                              &    \\
\hline
{ } 36    & 0468-0032035 &                  &                        &                              &    \\
\hline
{ } 37    & 0465-0031095 &                  &                        &                              &    \\
\hline
{ } 38    & 0468-0032036 &                  &                        &                              &    \\
\hline
{ } 39    &              &                  &                        &                              &    \\
\hline
{ } 40    & 0465-0031108 & 04034120-4329015 &   TYC7583-1149-1       &                              & foreground star   \\
\hline
{ } 41    & 0468-003204 &                   &                        &                              &     \\
\hline
{ } 42    &             &                   &                        &                              &     \\
\hline
{ } 43    & 0466-0031581 &                  &                        &                              &     \\
\hline  
{ } 44    &             &                   &                        &                              &     \\
\hline
{ } 45    &             &                   &                        &                              &    \\
\hline
{ } 46    &             &                   &                        &                              &    \\
\hline
{ } 47    &             &                   &                        &                              &    \\
\hline
{ } 48    &             &                   &                        &                              &    \\
\hline
{ } 49    &             &                   &                        &                              &    \\
\hline
{ } 50    &             &                   &                        &                              &    \\
\hline
{ } 51    &             &                   &                        &                              &    \\
\hline
{ } 52    &             &                  &                        &                              &    \\
\hline
{ } 53    &              &                  &                        &                              &    \\
\hline
{ } 54    & 0466-0031598 & 04035419-4320552 &      NGC1512           & ATCA 04 03 54.20 -43 20 56.2 &   \\
\hline
{ } 55    &             &                   &                        &                              &    \\
\hline
{ } 56    & 0467-0032138 &                  &                        &                              &    \\
\hline
{ } 57    &              &                  &                        &                              &    \\
\hline
{ } 58    & 0468-0032062 &                  &                        &                              &    \\
\hline
{ } 59    &              &                  &                        &                              &    \\
\hline
{ } 60    &              &                  &                        &                              &     \\
\hline
{ } 61    & 0465-0031140 & 04035632-4329041 &    MRSS~250-126782     &                              & galaxy   \\
\hline
{ } 62    & 0466-0031604 &                  &                        &                              &     \\
\hline
{ } 63    & 0466-0031610 & 04040093-4323175 &                        &                              & QSO or XRB \\
\hline
{ } 64    &              &                  &                        &                              &    \\
\hline
{ } 65    &              &                  &                        &                              &    \\
\hline
{ } 66    &              &                  &                        &                              &    \\
\hline
{ } 67    & 0466-0031624 &                  &                        & ATCA 04 04 08.82 -43 18 42.3 & background object    \\
\hline
{ } 68    &              &                  &                        &                              &    \\
\hline
{ } 69    &              &                  &                        &                              &    \\
\hline
{ } 70    &              &                  &                        &                              &    \\
\hline
{ } 71    &              &                  &                        & SUMSS\,J040412-432235 and ATCA source & background object \\
\hline
{ } 72    &              &                  &                        &                              &     \\
\hline
{ } 73    &              &                  &                        &                              &     \\
\hline
{ } 74    &              &                  &                        &                              &    \\
\hline
{ } 75    &              &                  &                        &                              &    \\
\hline
{ } 76    &              &                  &                        &                              &   \\
\hline
{ } 77    &              &                  &                        &                              &    \\
\hline
{ } 78    &              &                  &                        &                              &    \\
\hline
{ } 79    &              &                  &                        &                              &    \\
\hline
{ } 80    &              &                  &                        &                              &    \\
\hline
{ } 81    & 0465-0031199 &                  &                        &                              &    \\
\hline
{ } 82    & 0466-0031640 & 04042126-4322204 &                        &                              &        \\
\hline
{ } 83    &              &                  &                        & ATCA 04 04 22.25 -43 24 52.1 & background object   \\
\hline
{ } 84    &              &                  &                        &                              &    \\
\hline
{ } 85    &              &                  &                        &                              &    \\
\hline
{ } 86    &              &                  &                        &                              &      \\
\hline
{ } 87    &              &                  &                        &                              &    \\
\hline
{ } 88    &              &                  &                        &                              &    \\
\hline
{ } 89    & 0465-0031223 & 04043150-4326090 & TYC583-426-1           &                              & foreground star   \\
\hline
{ } 90    &              &                  &                        &                              &    \\
\hline
{ } 91    &              &                  &                        &                              &    \\
\hline
{ } 92    & 0467-0032209 &                  &                        &                              &         \\
\hline
{ } 93    & 0468-0032147 &                  &                        &                              &    \\
\hline
{ } 94    & 0465-0031250 & 04044352-4325227 &   MRSS 250-123273      & ATCA 04 04 43.66 -43 25 22.5 & foreground star   \\
\hline
{ } 95    &              &                  &                        &                              &     \\
\hline
{ } 96    &              &                  &                        &                              &     \\
\hline
{ } 97    &              &                  &                        &                              &    \\
\hline
{ } 98    &              &                  &                        & ATCA 04 04 46.77 -43 22 02.2 &  background object   \\
\hline
{ } 99    & 0466-0031693 &                  &                        &                              &     \\
\hline
100       &              &                  &                        &                              &     \\
\hline
101       & 0467-0032246 &                  &                        &                              &     \\
\hline
102       &              &                  &                        &                              &     \\
\hline
103       &              &                  &                        &                              &    \\
\hline
104       & 0466-0031732 &                  &                        &                              &    \\
\hline
105       & 0466-0031738 &                  &                        &                              &    \\
\hline
106       & 0466-0031740 &                  &                        &                              &    \\
\hline
\end{longtable}
\twocolumn

\end{appendix}

\end{document}